\documentclass[11pt,a4paper]{article}
\pdfoutput=1

\usepackage[nosort]{cite}

\usepackage[british]{babel} %dates in british format

\usepackage{epsfig,rotating}
\usepackage{amsmath,amssymb,amsthm}
\usepackage{amsfonts}
\usepackage{mathrsfs}
\usepackage{bbm}
\usepackage{bm}
\usepackage[normalem]{ulem}
\usepackage{mathtools}
\usepackage[all,knot]{xy}
\xyoption{arc}

\usepackage{tikz}
\usetikzlibrary{arrows}
\usetikzlibrary{patterns}
\usetikzlibrary{decorations.markings}

\usepackage{mathabx}

\usepackage{xcolor}
\usepackage[textwidth = 430 pt, textheight = 630 pt]{geometry}
\definecolor{MyDarkBlue}{rgb}{0.15,0.25,0.45}

\usepackage{enumerate}

\usepackage{hyperref}
\hypersetup{
    colorlinks=true,
    citecolor=MyDarkBlue,
    linkcolor=MyDarkBlue,
    urlcolor=MyDarkBlue,
    pdfauthor={Tommaso Macrelli, Christian S\"amann, Martin Wolf},
    pdftitle={Scattering Amplitude Recursion Relations in BV Quantisable Theories},
    pdfsubject={hep-th},
    breaklinks=true,
    linktocpage=true,
    hypertexnames=false
} 

\usepackage{axodraw2}

\let\fn\footnote
\renewcommand{\footnote}[1]{\linespread{1.1}\fn{#1}\linespread{1.29}}

\makeatletter
\renewcommand{\xRightarrow}[2][]{\ext@arrow 0359\Rightarrowfill@{#1}{#2}}
\renewcommand{\section}{\@startsection
    {section}{1}{\z@}{-3.5ex plus -1ex minus
        -.2ex}{2.3ex plus .2ex}{\bf\mathversion{bold} }}
\renewcommand{\subsection}{\@startsection{subsection}{2}{\z@}{-3.25ex
        plus -1ex minus
        -.2ex}{1.5ex plus .2ex}{\bf\mathversion{bold} }}
\renewcommand{\subsubsection}{\@startsection{subsubsection}{3}{-2.45ex}{-3.25ex
        \makeatother
        plus -1ex minus -.2ex}{1.5ex plus .2ex}{\it }}
\renewcommand{\thesection}{\arabic{section}}
\renewcommand{\thesubsection}{\arabic{section}.\arabic{subsection}}
\renewcommand{\@seccntformat}[1]{\@nameuse{the#1}.~~}

\renewcommand{\theequation}{\thesection.\arabic{equation}}
\makeatletter \@addtoreset{equation}{section}

\renewcommand*\l@section{\@dottedtocline{1}{0em}{2em}}
\renewcommand*\l@subsection{\@dottedtocline{2}{2em}{2.4em}}
\renewcommand*\l@subsubsection{\@dottedtocline{4}{3.8em}{3.7em}}

\renewcommand\tableofcontents{%
    \section*{\large\contentsname
        \@mkboth{%
            \MakeUppercase\contentsname}{\MakeUppercase\contentsname}}%
    {\baselineskip=15pt plus 2pt minus 1pt
        \@starttoc{toc}}%
}

\newcommand{\acknowledgements}{\section*{Acknowledgements}
    \addcontentsline{toc}{section}{Acknowledgements}}

\newcommand{\datamanagement}{\section*{Data Management}
    \addcontentsline{toc}{section}{Data Management}}

\newcommand{\appendices}{
    \section*{Appendix}\label{appendices}\setcounter{subsection}{0}
    \addcontentsline{toc}{section}{Appendix}
    \setcounter{equation}{0}
    \setcounter{thm}{0}
    \makeatletter
    \renewcommand{\theequation}{\Alph{subsection}.\arabic{equation}}
    \renewcommand{\thesubsection}{\Alph{subsection}}
    \renewcommand{\thethm}{\Alph{subsection}.\arabic{thm}}
    \@addtoreset{equation}{subsection}
    \@addtoreset{thm}{subsection}
    \makeatother
}

\renewcommand{\thethm}{\thesection.\arabic{thm}}

\def\periodb#1{\setbox0=\hbox{$#1$}#1\hskip-\wd0\hbox to\wd0{-}}

\newcommand{\im}{\mathrm{im}}   			% identity map/matrix
\newcommand{\id}{\mathrm{id}}   			% identity map/matrix
\newcommand{\CA}{\mathcal{A}}    			% cal-letters

\newcommand{\CCC}{\mathscr{C}}

\newcommand{\CCS}{\mathscr{S}}

\newcommand{\Sh}{\mathrm{Sh}}     			% trace

\newcommand{\CO}{\mathcal{O}}

\newcommand{\MM}{\FR^{1,3}}     			% set of natural numbers

\newcommand{\frg}{\mathfrak{g}}				% frak-letters
\newcommand{\frl}{\mathfrak{l}}				% frak-letters
\newcommand{\fra}{\mathfrak{a}}				% frak-letters

\newcommand{\frF}{\mathfrak{F}}

\newcommand{\FR}{\mathbbm{R}}     			% field of real numbers
\newcommand{\NN}{\mathbbm{N}}     			% set of natural numbers
\newcommand{\RZ}{\mathbbm{Z}}     			% ring of integers
\newcommand{\dd}{\mathrm{d}}     			% total differential
\newcommand{\dpar}{\partial}     			% partial differential
\newcommand{\embd}{{\hookrightarrow}}     		% embedded
\newcommand{\de}{\mathrm{e}}     			% Euler's number
\newcommand{\di}{\mathrm{i}}     			% imaginary unit
\newcommand{\eps}{{\varepsilon}}			% antisymmetric tensors
\newcommand{\eand}{{~~~\mbox{and}~~~}}     		% and etc. in equations
\newcommand{\ewith}{{~~~\mbox{with}~~~}}
\newcommand{\efor}{{~~~\mbox{for}~~~}}

\newcommand{\tr}{\,\mathrm{tr}\,}     			% trace
\newcommand{\ad}{\mathrm{ad}}     			% adjoint action
\newcommand{\sL}{\sfL}

\newcommand{\remark}[1]{}     				% remark
\newcommand{\myxymatrix}[1]{\vcenter{\vbox{\xymatrix{#1}}}}
\def\tyng(#1){\hbox{\tiny$\yng(#1)$}}			% small Young diagram
\def\tyoung(#1){\hbox{\tiny$\young(#1)$}}			% small Young diagram
\newcommand{\sfh}{\mathsf{h}}

\newcommand{\sfp}{\mathsf{p}}
\newcommand{\sfe}{\mathsf{e}}

\newcommand{\sfL}{\mathsf{L}}

\newcommand{\ldsb}{[\![}
\newcommand{\rdsb}{]\!]}

\begin{document}
    \begin{titlepage}
        
        \setcounter{page}{0}
        \renewcommand{\thefootnote}{\fnsymbol{footnote}}
        
        \begin{flushright}
            DMUS--MP--19/04\\
            EMPG--19--09
        \end{flushright}
        
        \begin{center}
            
            \vspace{2cm}

            {\LARGE\textbf{\mathversion{bold}Scattering Amplitude Recursion Relations\\  in BV Quantisable Theories}\par}
            
            \vspace{1cm}
            
            {\large
                Tommaso Macrelli$^{a}$, Christian S\"amann$^{b}$, and Martin Wolf$^{\,a}$
                \footnote{{\it E-mail addresses:\/}
                    \href{mailto:t.macrelli@hw.ac.uk}{\ttfamily t.macrelli@surrey.ac.uk}, 
                    \href{mailto:c.saemann@hw.ac.uk}{\ttfamily c.saemann@hw.ac.uk}, 
                    \href{mailto:m.wolf@surrey.ac.uk}{\ttfamily m.wolf@surrey.ac.uk}
                }}
            
            \vspace{.3cm}
            
            {\it
                $^a$
                Department of Mathematics,
                University of Surrey\\
                Guildford GU2 7XH, United Kingdom\\[.3cm]
                
                $^b$ Maxwell Institute for Mathematical Sciences\\
                Department of Mathematics,
                Heriot--Watt University\\
                Edinburgh EH14 4AS, United Kingdom
            }
            
            \vspace{2cm}
            
            {\bf Abstract}
        \end{center}
        \vspace{-.5cm}
        \begin{quote}
            Tree-level scattering amplitudes in Yang--Mills theory satisfy a recursion relation due to Berends and Giele which yields e.g.~the famous Parke--Taylor formula for MHV amplitudes. We show that the origin of this recursion relation becomes clear in the BV formalism, which encodes a field theory in an $L_\infty$-algebra. The recursion relation is obtained in the transition to a smallest representative in the quasi-isomorphism class of that $L_\infty$-algebra, known as a minimal model. In fact, the quasi-isomorphism contains all the information about the scattering theory. As we explain, the computation of such a minimal model is readily performed in any BV quantisable theory, which, in turn, produces recursion relations for its tree-level scattering amplitudes.
            
            \vfill\noindent 21st October 2020%\today
            
        \end{quote}
        
        \setcounter{footnote}{0}\renewcommand{\thefootnote}{\arabic{thefootnote}}
        
    \end{titlepage}
    
    \tableofcontents
    \bigskip
    \bigskip
    \hrule
    \bigskip
    \bigskip

    \section{Introduction and results}
    
    Whilst string theory has not yet fulfilled its initial promise of a complete and unified description of nature, it has certainly become a successful way of thinking about quantum field theories. Here, we would like to adopt a perspective which has its origin in string field theory and which was suggested e.g.~in~\cite{Kajiura:2001ng,Kajiura:0306332,Jurco:2018sby,Jurco:2019bvp}. The structures of the Hilbert spaces of string field theories are encoded in homotopy algebras, and in the case of closed string field theory in terms of {\em $L_\infty$-algebras}~\cite{Zwiebach:1992ie}. The relevant classical action is simply  the canonical action associated with an $L_\infty$-algebra and which is known as the {\em homotopy Maurer--Cartan action}. Homotopy Maurer--Cartan theory can be thought of as a vast generalisation of Chern--Simons theory. One might hope that this rich structure is somehow reflected in ordinary field theory, which one could then exploit, e.g.~in often cumbersome computations of scattering amplitudes. As we shall see, this is indeed the case and the place to look for $L_\infty$-algebras is the Batalin--Vilkovisky (BV) formalism~\cite{Batalin:1977pb,Batalin:1981jr,Batalin:1984jr,Batalin:1984ss,Batalin:1985qj}, see also e.g.~\cite{Vasiliev:2005zu,Barnich:2005ru} and~\cite{Zeitlin:2007fp} for related, earlier considerations.
    
    In this paper, we shall combine the following three facts, which should be familiar to any expert on BV quantisation and which are explained in detail e.g.~in~\cite{Jurco:2018sby,Jurco:2019bvp}:
    \begin{enumerate}[i)]
        \setlength{\itemsep}{-2pt}
        \item The BV formalism assigns to any field theory it can treat an $L_\infty$-algebra describing its symmetries, field contents, equations of motion, and Noether currents;
        \item Quasi-isomorphic $L_\infty$-algebras describe physically equivalent field theories~\cite{Jurco:2018sby} (see also~\cite{Barnich:2004cr});
        \item Any $L_\infty$-algebra comes with a {\em minimal model} which is a smallest representative in its quasi-isomorphism class and whose propagators vanish. 
    \end{enumerate}
    Given an $L_\infty$-algebra of a classical field theory, we are thus led to conclude that the $n$-point vertices of the field theory described by its minimal model should be the tree-level scattering amplitudes of the original field theory.
    
    An obvious candidate for investigating the validity of our conclusion is the famous Parke--Taylor formula, which describes a huge simplification in adding up the Feynman diagrams contributing to maximally helicity violating (MHV) gluon scattering amplitudes. After its conjecture in~\cite{Parke:1986gb}, this formula was proved by Berends and Giele in~\cite{Berends:1987me} using a recursion relation for particular currents. We shall show that this recursion relation is nothing but the explicit formula for computing a minimal model in the concrete case of Yang--Mills theory.
    
    We begin with a very concise review of $L_\infty$-algebras, quasi-isomorphisms, minimal models, and their appearance in the BV formalism in~Section~\ref{sec:L_infty_algebras}. We then discuss our formalism in~Section~\ref{sec:scalar_FT} for the example of scalar field theory in which the relevant structures and their interpretation become very obvious. Full Yang--Mills theory and the gluon current recursion relations are then discussed in Section~\ref{sec:YM}. We explain that the Berends--Giele recursion relations for tree-level scattering amplitudes in Yang--Mills theory arise as recursion relations of an underlying quasi-isomorphism of $L_\infty$-algebras. 
    
    Let us also summarise a number of general important observations that follow from our constructions. Any BV quantisable field theory gives rise to an $L_\infty$-algebra $\sL$. By the strictification theorem for $L_\infty$-algebras, there is a strict $L_\infty$-algebra $\tilde \sL$ that is quasi-isomorphic to $\sL$. In other words, any BV quantisable field theory can be cast into an equivalent field theory which has only propagators and cubic interaction vertices. Notice, however, whilst always guaranteed in theory, finding the explicit strict version of a theory might be difficult in practice. Furthermore, the minimal model $\sL^\circ$ of the $L_\infty$-algebra $\sL$ can always be constructed recursively, and this construction involves a particular map, taking the role of a contracting homotopy, which can be chosen to include the Feynman propagator of the theory. This explains that the minimal model describes the tree-level scattering amplitudes of the original field theory. Other choices of the contracting homotopy, however, are possible, and we expect interesting results for perturbation theory to emerge also from those.
    
    There are clearly many avenues for further study of the structures we discussed. Most important is certainly the development of the full quantum picture, going beyond the tree level. We note that besides deeper insights into the symmetries and structures of Feynman diagrams, this research may also lead to a formulation of quantum field theory in purely algebraic terms, which should be much more accessible to mathematicians than the standard textbook presentation.
    
    Whilst finishing this paper, we received the announcement of the forthcoming paper~\cite{Arvanitakis:2019ald} by Alexandros Arvanitakis in which he also discusses the S-matrix in terms of minimal models. At the same time, we became aware of the preprint~\cite{Nutzi:2018vkl} in which a mathematical explanation of the on-shell Britto--Cachazo--Feng--Witten (BCFW) recursion relations~\cite{Britto:2004ap,Britto:2005fq} via minimal models of $L_\infty$-algebras was given. Contrary to our general constructions in which the $L_\infty$-algebra of a field theory always arises from the BV formalism, the discussion in~\cite{Nutzi:2018vkl} relies on the explicit strict models of the $L_\infty$-algebras for the considered field theories. We also note that the idea of obtaining scattering amplitudes from minimal models is not new and can be traced back, at least, to~\cite{Kajiura:2001ng,Kajiura:0306332}. The latter references also developed much of the technology we are using in the following.

    \section{\texorpdfstring{$L_\infty$}{Linfinity}-algebras of field theories}\label{sec:L_infty_algebras}
    
    $L_\infty$-algebras are generalisations of Lie algebras to differential graded Lie algebras and beyond. We will review the relevant definitions and refer the interested reader to~\cite{Jurco:2018sby,Jurco:2019bvp} for all the details.
    
    \subsection{\texorpdfstring{$L_\infty$}{Linfinity}-algebras and quasi-isomorphisms}\label{ssec:L_infty_algebras}
    
    \paragraph{\mathversion{bold}$L_\infty$-algebras.}
    To begin with, let $\sL\coloneqq\bigoplus_{k\in\RZ}\sL_k$ be a $\RZ$-graded vector space. Elements of $\sL_k$ are said to be homogeneous and of degree~$k$, and we shall denote the degree of a homogeneous element $\ell\in\sfL$ by $|\ell|_\sfL\in\RZ$. Suppose there is a differential $\mu_1:\sL\to\sL$ of degree~$1$. This allows us to consider the chain complex \begin{subequations}
        \begin{equation}\label{eq:chainComplex}
            \cdots\ \xrightarrow{\phantom{~~\mu_1~~}}\ \sL_{-1}\ \xrightarrow{~~\mu_1~~}\ \sL_0\ \xrightarrow{~~\mu_1~~}\ \sL_1\ \xrightarrow{\phantom{~~\mu_1~~}}\ \cdots
        \end{equation}
        Next, we equip this complex with products $\mu_i:\sL\times\cdots\times\sL\to\sL$ of degree~$2-i$ for $i\in\NN$ which are $i$-linear and totally graded antisymmetric and subject to the \emph{higher} or \emph{homotopy Jacobi identity}\footnote{There are many possible sign conventions here; we believe we chose the convention with the least amount of overhead of signs for the BV formalism.}
        \begin{equation}\label{eq:homotopyJacobi}
            \sum_{j+k=i}\sum_{\sigma\in \Sh(j;i) }\chi(\sigma;\ell_1,\ldots,\ell_{i})(-1)^{k}\mu_{k+1}(\mu_j(\ell_{\sigma(1)},\ldots,\ell_{\sigma(j)}),\ell_{\sigma(j+1)},\ldots,\ell_{\sigma(i)})\ =\ 0
        \end{equation}
        for $\ell_1,\ldots,\ell_{i}\in \sfL$. The sum over $\sigma$ is taken over all $(j;i)$ \emph{shuffles} which consist of permutations $\sigma$ of $\{1,\ldots,i\}$ such that the first $j$ and the last $i-j$ images of $\sigma$ are ordered: $\sigma(1)<\cdots<\sigma(j)$ and $\sigma(j+1)<\cdots<\sigma(i)$. Furthermore, the sign $\chi(\sigma;\ell_1,\ldots,\ell_i)$ is called the \emph{graded Koszul sign} and defined by means of
        \begin{equation}
            \ell_1\wedge \ldots \wedge \ell_i\ =\ \chi(\sigma;\ell_1,\ldots,\ell_i)\,\ell_{\sigma(1)}\wedge \ldots \wedge \ell_{\sigma(i)}~.
        \end{equation}
    \end{subequations}
    Specifically, when $i=1$, the homotopy Jacobi identity~\eqref{eq:homotopyJacobi} just says that $\mu_1$ is a differential. For $i=2$, it says that $\mu_1$ is a derivation with respect to $\mu_2$ and for $i=3$, it says that the binary product $\mu_2$ satisfies a generalisation of the standard Jacobi identity, and so on.
    
    A $\RZ$-graded vector space with such products $\mu_i$ is called an \emph{$L_\infty$-algebra}~\cite{Stasheff:1992bb,Lada:1992wc,Lada:1994mn}. Particular examples of $L_\infty$-algebras include the {\em trivial $L_\infty$-algebra} $\sL=\bigoplus_{k\in\RZ}\sL_k$ with $\sL=\{0\}$, ordinary Lie algebras with $\sL=\sL_0$ and the only non-vanishing product being $\mu_2$, as well as differential graded Lie algebras with general $\sL$ for which $\mu_i=0$ for $i\geq 3$. The latter are also called {\em strict $L_\infty$-algebras}.
    
    \paragraph{\mathversion{bold}$L_\infty$-morphisms.}
    Morphisms between Lie algebras are maps preserving the Lie bracket. The higher categorical nature of $L_\infty$-algebras now leads to a vast generalisation of Lie algebra morphisms to \emph{$L_\infty$-morphisms}. Explicitly, an $L_\infty$-morphism $\phi:(\sfL,\mu_i)\to (\sfL',\mu'_i)$ between two $L_\infty$-algebras $(\sfL,\mu_i)$ and $(\sfL',\mu'_i)$ is a collection of $i$-linear totally graded antisymmetric maps $\phi_i:\sL\times\cdots\times\sL\to\sL'$ of degree~$1-i$ such that
    \begin{subequations}\label{eq:L_infty_morphism}
        \begin{equation}
            \begin{aligned}
                &\sum_{j+k=i}\sum_{\sigma\in \Sh(j;i)}~(-1)^{k}\chi(\sigma;\ell_1,\ldots,\ell_i)\phi_{k+1}(\mu_j(\ell_{\sigma(1)},\dots,\ell_{\sigma(j)}),\ell_{\sigma(j+1)},\dots ,\ell_{\sigma(i)})\\
                \ &=\ \sum_{j=1}^i\frac{1}{j!} \sum_{k_1+\cdots+k_j=i}\sum_{\sigma\in{\rm Sh}(k_1,\ldots,k_{j-1};i)}\chi(\sigma;\ell_1,\ldots,\ell_i)\zeta(\sigma;\ell_1,\ldots,\ell_i)\,\times\\
                &\kern1cm\times \mu'_j\Big(\phi_{k_1}\big(\ell_{\sigma(1)},\ldots,\ell_{\sigma(k_1)}\big),\ldots,\phi_{k_j}\big(\ell_{\sigma(k_1+\cdots+k_{j-1}+1)},\ldots,\ell_{\sigma(i)}\big)\Big)
            \end{aligned}
        \end{equation}
        with $\chi(\sigma;\ell_1,\ldots,\ell_i)$ the Koszul sign and $\zeta(\sigma;\ell_1,\ldots,\ell_i)$ given by
        \begin{equation}\label{eq:zeta-sign}
            \zeta(\sigma;\ell_1,\ldots,\ell_i)\ \coloneqq\ (-1)^{\sum_{1\leq m<n\leq j}k_mk_n+\sum_{m=1}^{j-1}k_m(j-m)+\sum_{m=2}^j(1-k_m)\sum_{k=1}^{k_1+\cdots+k_{m-1}}|\ell_{\sigma(k)}|_\sL}~.
        \end{equation}
    \end{subequations}
    Note that for Lie algebras, this definition just reduces to the standard definition of a Lie algebra morphism. 
    
    Since $\mu_1$ is a differential, we may study the cohomology ring $H^\bullet_{\mu_1}(\sL)$ of the chain complex~\eqref{eq:chainComplex}. If the map $\phi_1$ of an $L_\infty$-morphism $\phi:(\sfL,\mu_i)\to (\sfL',\mu'_i)$ induces an isomorphism on the cohomology rings $H^\bullet_{\mu_1}(\sL)\cong H^\bullet_{\mu_1}(\sL')$, then $\phi$ is called a \emph{quasi-isomorphism}, generalising quasi-isomorphisms of chain complexes. Quasi-isomorphisms are, in most cases, the appropriate notion of isomorphisms for $L_\infty$-algebras.
    
    \paragraph{Strictification theorem.} General strictification theorems for homotopy algebras~\cite{igor1995,Berger:0512576} specialise to $L_\infty$-algebras and state that any $L_\infty$-algebra is quasi-isomorphic to a strict $L_\infty$-algebra. The latter is then called a {\em strict model} for the former. Recall that a strict $L_\infty$-algebra is a differential graded Lie algebra because only the differential and the binary product are non-vanishing. 
    
    We shall see explicit examples of strictifications of $L_\infty$-algebras in Sections~\ref{ssec:scalar_strict} and~\ref{ssec:YM_strict}. In practice, however, it often turns out that the transition to the strict model of an $L_\infty$-algebra is either hard to begin with or not very convenient and too restrictive for further computations.
    
    \paragraph{Minimal model theorem.} A companion theorem to the above one is the {\em minimal model theorem}\footnote{See~\cite{kadeishvili1982algebraic,Kajiura:0306332} for the corresponding statement for $A_\infty$-algebras.}: any $L_\infty$-algebra $(\sfL,\mu_i)$ is quasi-isomorphic to an $L_\infty$-algebra $(\sfL^\circ,\mu^\circ_i)$ for which $\mu^\circ_1=0$. The latter is indeed a {\em minimal model}, i.e.~a minimal representative of the quasi-isomorphism class of $\sL$, since the graded vector space underlying $\sfL^\circ$ is its own cohomology ring $H^\bullet_{\mu_1}(\sfL)$. We note that minimal models are unique up to $L_\infty$-isomorphisms (i.e.~$L_\infty$-morphisms $\phi$ with $\phi_1$ invertible).
    
    The construction of a minimal model for an $L_\infty$-algebra $\sL$ means to compute the $L_\infty$-structure given by brackets $\mu_i^\circ$ on the cohomology ring $H^\bullet_{\mu_1}(\sfL)=:\sfL^\circ$. Since this will be central to our discussion, we give some more details. We start from a choice of projection $\sfp:\sL\twoheadrightarrow \sL^\circ$ together with an embedding $\sfe:\sL^\circ\,\embd\, \sL$ which are both chain maps of degree~0 and satisfy $\sfp\circ \sfe=\id_{\sL^\circ}$. Then we always have a degree~$-1$ chain map $h:\sL\rightarrow \sL$ satisfying
    \begin{equation}\label{eq:contracting_homotopy}
        \id_{\sL}-\sfe\circ \sfp\ =\ h\circ \mu_1+\mu_1\circ h~.
    \end{equation}
    Such a map $h$ is called a \emph{contracting homotopy}, and we summarise this pictorially as 
    \begin{equation}
        \myxymatrix{\ar@(dl,ul)[]^h \sfL~\ar@<+2pt>@{->>}[rr]^{\kern-20pt \sfp} & & ~H^\bullet_{\mu_1}(\sfL) \ar@<+2pt>@{^(->}[ll]^{\kern-20pt \sfe}}.
    \end{equation}
    Evidently, since $(\sL,\mu_1)$ is a complex and $\sfp$ and $\sfe$ are chain maps, we have 
    \begin{equation}\label{eq:contracting_identities}
        \mu_1\circ \mu_1\ =\ 0~,~~~\mu_1\circ \sfe\ =\ 0~,\eand\sfp\circ \mu_1\ =\ 0~.
    \end{equation}
    Upon combining this with~\eqref{eq:contracting_homotopy}, we obtain
    \begin{equation}
        \mu_1\ =\ \mu_1\circ h \circ \mu_1~.
    \end{equation}
    The map $\sfe\circ \sfp$ in~\eqref{eq:contracting_homotopy} is a projector onto a subspace of $\sL$, however, the maps $h\circ \mu_1$  and $\mu_1\circ h$ on the right-hand side are not, in general. We can always rectify this by redefining the contracting homotopy according to
    \begin{equation}\label{eq:improvement_h}
        \sfh\ \coloneqq\ h-h\circ h\circ \mu_1-h\circ \sfe\circ \sfp~.
    \end{equation}
    Indeed, with the help of~\eqref{eq:contracting_homotopy} and~\eqref{eq:contracting_identities}, one may check that $\sfh$ satisfies
    \begin{equation}\label{eq:contracting_homotopy_htilde}
        \begin{gathered}
            \id_{\sL}-\sfe\circ \sfp\ =\ \sfh\circ \mu_1+\mu_1\circ \sfh~,\\
            \sfh\ =\ \sfh-\sfh\circ \sfh\circ \mu_1-\sfh\circ \sfe\circ \sfp~,\\
            \mu_1\ =\ \mu_1\circ \sfh \circ \mu_1~,~~~\sfh\ =\ \sfh\circ \mu_1\circ \sfh~,~~~\sfh\circ \sfh\ =\ 0~.
        \end{gathered} 
    \end{equation}
    In particular, $\sfh$ is again a contracting homotopy. Moreover, we now have a decomposition\footnote{The subscripts are borrowed from the Hodge decomposition of a differential form into harmonic, exact, and co-exact parts, see~\cite[Section 5.2]{Jurco:2018sby} for the corresponding formulas.}
    \begin{equation}\label{eq:Hodge-Kodaira-decomposition}
        \begin{gathered}
            \sL\ \cong\ \sL_{\rm harm}\oplus \sL_{\rm ex}\oplus \sL_{\rm coex}~,\\
            \sL_{\rm harm}\ \coloneqq\ \im(\sfe\circ \sfp)~,~~~\sL_{\rm ex}\ \coloneqq\ \im(\mu_1 \circ \sfh)~,~~~\sL_{\rm coex}\ \coloneqq\ \im(\sfh\circ \mu_1)
        \end{gathered}
    \end{equation}
    with $\sL_{\rm harm}\cong \sL^\circ$. This is known as the abstract \emph{Hodge--Kodaira decomposition}. It is rather straightforward to verify that 
    \begin{equation}
        \begin{gathered}
            \im(\sfe)\ \cong\ \sL_{\rm harm}~,~~~\im(\mu_1)\ \cong\ \sL_{\rm ex}~,~~~\im(\sfh)\ \cong\ \sL_{\rm coex}~,\\ 
            \ker(\sfp)\ \cong\ \sL_{\rm ex}\oplus \sL_{\rm coex}~,~~~\ker(\mu_1)\ \cong\ \sL_{\rm harm}\oplus \sL_{\rm ex}~,~~~\ker(\sfh)\ \cong\ \sL_{\rm harm}\oplus \sL_{\rm coex}~.
        \end{gathered}
    \end{equation}
    
    The quasi-isomorphism between $(\sfL,\mu_i)$ and $(\sfL^\circ,\mu_i^\circ)$ is now determined by the maps $\phi_i:\sfL^\circ\times\cdots\times\sfL^\circ\to\sfL$ which are constructed recursively as~\cite{Kajiura:0306332}
    \begin{subequations}\label{eq:minimal_model}
        \begin{equation}\label{eq:quasi_iso_phi}
            \begin{aligned}
                \phi_1(\ell^\circ_1)\ &\coloneqq\ \sfe(\ell^\circ_1)~,\\
                \phi_2(\ell^\circ_1,\ell^\circ_2)\ &\coloneqq\ - (\sfh\circ\mu_2)(\phi_1(\ell^\circ_1),\phi_1(\ell^\circ_2))~,\\
                &~~\vdots\\
                \phi_i(\ell^\circ_1,\ldots,\ell^\circ_i)\ &\coloneqq\ -\sum_{j=2}^i\frac{1}{j!} \sum_{k_1+\cdots+k_j=i}\sum_{\sigma\in{\rm Sh}(k_1,\ldots,k_{j-1};i)}\chi(\sigma;\ell^\circ_1,\ldots,\ell^\circ_i)\zeta(\sigma;\ell^\circ_1,\ldots,\ell^\circ_i)\,\times\\
                &\kern1cm\times (\sfh\circ\mu_j)\big(\phi_{k_1}\big(\ell^\circ_{\sigma(1)},\ldots,\ell^\circ_{\sigma(k_1)}\big),\ldots,\phi_{k_j}\big(\ell^\circ_{\sigma(k_1+\cdots+k_{j-1}+1)},\ldots,\ell^\circ_{\sigma(i)}\big)\big)
            \end{aligned}
        \end{equation}
        and likewise, the brackets $\mu^\circ_i:\sfL^\circ\times\cdots\times\sfL^\circ\to\sfL^\circ$ are given by~\cite{Kajiura:0306332} 
        \begin{equation}
            \begin{aligned}
                \mu^\circ_1(\ell^\circ_1)\ &\coloneqq\ 0~,\\
                \mu^\circ_2(\ell^\circ_1,\ell^\circ_2)\ &\coloneqq\ (\sfp\circ\mu_2)(\phi_1(\ell^\circ_1),\phi_1(\ell^\circ_2))~,\\
                &~~\vdots\\
                \mu^\circ_i(\ell^\circ_1,\ldots,\ell^\circ_i)\ &\coloneqq\ \sum_{j=2}^i\frac{1}{j!} \sum_{k_1+\cdots+k_j=i}\sum_{\sigma\in{\rm Sh}(k_1,\ldots,k_{j-1};i)}\chi(\sigma;\ell^\circ_1,\ldots,\ell^\circ_i)\zeta(\sigma;\ell^\circ_1,\ldots,\ell^\circ_i)\,\times\\
                &\kern1cm\times (\sfp\circ\mu_j)\big(\phi_{k_1}\big(\ell^\circ_{\sigma(1)},\ldots,\ell^\circ_{\sigma(k_1)}\big),\ldots,\phi_{k_j}\big(\ell^\circ_{\sigma(k_1+\cdots+k_{j-1}+1)},\ldots,\ell^\circ_{\sigma(i)}\big)\big)~,
            \end{aligned}
        \end{equation}
    \end{subequations}
    where $\ell^\circ_1,\ldots,\ell^\circ_i\in\sfL^\circ$. Here, $\chi(\sigma;\ell^\circ_1,\ldots,\ell^\circ_i)$ is the Koszul sign and $\zeta(\sigma;\ell^\circ_1,\ldots,\ell^\circ_i)$ the sign factor introduced in~\eqref{eq:zeta-sign}. We shall provide a proof of these formulas in Appendix~\ref{app:proof_minimal_model}, explaining the derivation in~\cite{Kajiura:0306332} in some more detail.
    
    \paragraph{\mathversion{bold}Cyclic $L_\infty$-algebras.}
    The appropriate notion of a metric (or indefinite inner product) on an $L_\infty$-algebra is the following one. A {\em cyclic structure} on an $L_\infty$-algebra $(\sfL,\mu_i)$ is a non-degenerate bilinear graded symmetric pairing $\langle-,-\rangle_\sL:\sfL\times\sfL\to\FR$ of degree~$k$ which is cyclic in the sense of
    \begin{equation}\label{eq:cyclicity}
        \langle\ell_1,\mu_i(\ell_2,\ldots,\ell_{i+1})\rangle_\sL\ =\ (-1)^{i+i(|\ell_1|_\sL+|\ell_{i+1}|_\sL)+|\ell_{i+1}|_\sL\sum_{j=1}^{i}|\ell_j|_\sL}\langle\ell_{i+1},\mu_i(\ell_1,\ldots,\ell_{i})\rangle_\sL
    \end{equation}
    for $\ell_1,\ldots,\ell_{i+1}\in\sfL$. An $L_\infty$-algebra quipped with an inner product is called a \emph{cyclic $L_\infty$-algebra}.
    
    We can extend morphisms of $L_\infty$-algebras $\phi:(\sfL,\mu_i)\to(\sfL',\mu'_i)$ to morphisms of cyclic $L_\infty$-algebras if we additionally require
    \begin{subequations}\label{eq:compatibility_cyclic_morphism}
        \begin{equation}
            \langle \phi_1(\ell_1),\phi_1(\ell_2)\rangle_{\sL'}\ =\ \langle \ell_1,\ell_2\rangle_\sL
        \end{equation}
        and for all $i\geq 3$ and $\ell_1,\dots, \ell_i\in \sL$,
        \begin{equation}
            \sum_{\substack{j+k=i\\j,k\geq 1}}\langle \phi_j(\ell_1,\dots,\ell_j),\phi_k(\ell_{j+1},\dots,\ell_{j+k}))\rangle_{\sL'}\ =\ 0~.
        \end{equation}
    \end{subequations}
    Likewise, the strictification and minimal model theorems extend to cyclic $L_\infty$-algebras.
    
    \subsection{Homotopy Maurer--Cartan theory}
    
    The BV formalism can be seen as a reformulation of a Lagrangian field theory as a homotopy Maurer--Cartan theory, which is a  generalised form of a Chern--Simons theory. In the following, we concisely recall the basic facts and refer again to~\cite{Jurco:2018sby,Jurco:2019bvp} for more details.
    
    \paragraph{Homotopy Maurer--Cartan equation.}
    Given an $L_\infty$-algebra $(\sfL,\mu_i)$, we call an element of degree~1, $a\in\sfL_1$,  a \emph{gauge potential} and define its \emph{curvature} by 
    \begin{equation}\label{eq:curvature}
        f\ \coloneqq\ \sum_{i\geq 1}\frac{1}{i!}\mu_i(a,\ldots,a)~\in~\sL_2.
    \end{equation}
    Due to the higher Jacobi identities~\eqref{eq:homotopyJacobi}, the curvature $f$ obeys the \emph{Bianchi identity}
    \begin{equation}\label{eq:Bianchi}
        \sum_{i\geq0}\frac{1}{i!}\mu_{i+1}(a,\ldots,a,f)\ =\ 0~.
    \end{equation}
    
    Infinitesimal \emph{gauge transformations} are mediated by degree~0 elements $c_0\in\sfL_0$ and are given by $a\mapsto a+\delta_{c_0}a$ with
    \begin{equation}\label{eq:gauge_trafo}
        \delta_{c_0}a\ \coloneqq\ \sum_{i\geq0} \frac{1}{i!}\mu_{i+1}(a,\ldots,a, c_0)\quad\Longrightarrow\quad \delta_{c_0}f\ =\ \sum_{i\geq 0}\frac{1}{i!}\mu_{i+2}(a,\ldots,a, f,c_0)~.
    \end{equation}
    Using the higher Jacobi identities~\eqref{eq:homotopyJacobi}, one may show that 
    \begin{subequations}
        \begin{equation}
            [\delta_{ c_0},\delta_{ c'_0}]a\ =\ \delta_{c''_0}a+\sum_{i\geq0}\frac{1}{i!}\mu_{i+3}(a,\ldots,a, f,c_0, c'_0)~,
        \end{equation}
        where
        \begin{equation}
            c''_0\ \coloneqq\ \sum_{i\geq0}\frac{1}{i!}\mu_{i+2}(a,\ldots,a, c_0, c'_0)~.
        \end{equation}
    \end{subequations}
    Thus, the gauge transformations always close for strict $L_\infty$-algebras, for which only the differential and the 2-product are non-trivial. In the general case, however, a restriction of the gauge potential is required to ensure closure, and a sufficient condition is 
    \begin{equation}
        f\ =\ 0~.
    \end{equation}
    This equation, which describes an abstract form of flatness of the gauge potential, is known as the \emph{homotopy Maurer--Cartan  equation}. Gauge potentials $a\in\sfL_1$ satisfying this equation are called \emph{Maurer--Cartan elements}.
    
    It is important to stress that the gauge parameters $c_0\in\sfL_0$ may enjoy gauge freedom themselves which is mediated by \emph{next-to-lowest} gauge parameters $c_{-1}\in\sfL_{-1}$ of degree~$-1$. In turn, the next-to-lowest gauge parameters $c_{-1}\in\sfL_{-1}$ enjoy gauge freedom that is mediated by \emph{next-to-next-to-lowest} gauge parameters $c_{-2}\in\sfL_{-2}$ for degree~$-2$, and so on. These are known as the \emph{higher gauge transformations} and they are given by
    \begin{equation}
        \delta_{c_{-k-1}}  c_{-k}\ \coloneqq\ \sum_{i\geq0} \frac{1}{i!}\mu_{i+1}(a,\ldots,a, c_{-k-1})
    \end{equation}
    with $c_{-k}\in\sfL_{-k}$ of degree~$-k$. Note that $f=0$ is also a sufficient condition for the higher gauge transformations to close.
    
    \paragraph{Homotopy Maurer--Cartan action.}
    The homotopy Maurer--Cartan equation is variational whenever $(\sfL,\mu_i,\langle-,-\rangle)$ is a cyclic $L_\infty$-algebra with an inner product $\langle-,-\rangle$ of degree~$-3$. Indeed, the gauge invariant action functional
    \begin{equation}\label{eq:hMC_action}
        S_{\rm MC}\ \coloneqq\ \sum_{i\geq1} \frac{1}{(i+1)!}\,\langle a,\mu_i(a,\ldots,a)\rangle~,
    \end{equation}
    known as the {\em homotopy Maurer--Cartan action}, has the homotopy Maurer--Cartan equation as its stationary locus.
    
    \paragraph{\mathversion{bold}Homotopy Maurer--Cartan elements and $L_\infty$-morphisms.}
    Let us now briefly explain how Maurer--Cartan elements transform under  $L_\infty$-morphisms. For any $L_\infty$-morphism $\phi$ between two $L_\infty$-algebras $(\sfL,\mu_i)$ and  $(\sfL',\mu'_i)$, there is a natural morphism of gauge potentials,
    \begin{equation}\label{eq:map_hMC_to_hMC}
        a\ \mapsto\ a'\ \coloneqq\ \sum_{i\geq 1} \frac{1}{i!}\phi_i(a,\dots,a)\quad\Longrightarrow\quad f\ \mapsto\ f'\ =\ \sum_{i\geq 0}\frac{1}{i!}\phi_{i+1}(a,\ldots,a,f)~,
    \end{equation}
    which thus maps Maurer--Cartan elements to Maurer--Cartan elements.
    
    Furthermore, a gauge transformation $a\mapsto a+\delta_{c_0}a$ with gauge parameter $c_0\in\sfL_0$ of a Maurer--Cartan element $a\in\sfL_1$ is transformed under an $L_\infty$-morphism to  $a'\mapsto a'+\delta_{c'_0}a'$ with $a'\in\sfL'_1$ given by~\eqref{eq:map_hMC_to_hMC} and 
    \begin{equation}
        c_0\ \mapsto\ c_0'\ \coloneqq\ \sum_{i\geq 0} \frac{1}{i!}\phi_{i+1}(a,\dots,a, c_0)~.
    \end{equation}
    Consequently, gauge equivalence classes of Maurer--Cartan elements are mapped to gauge equivalence classes of Maurer--Cartan elements. Note that whenever $\phi$ is a quasi-isomor\-phism, the moduli space of Maurer--Cartan elements for $(\sfL,\mu_i)$  (that is, the space of solutions to the homotopy Maurer--Cartan equation modulo gauge transformations) is isomorphic to the moduli space of Maurer--Cartan elements for $(\sfL',\mu'_i)$.
    
    \subsection{Field theory and underlying \texorpdfstring{$L_\infty$}{Linfinity}-structures}\label{ssec:FT_from_L_infty}
    
    \paragraph{Classical observables.}
    The most general approach to the quantisation of gauge theories is certainly the BV formalism. To prepare the field theory for quantisation, the BV formalism constructs a modern description of the space of classical observables, which are the functionals on the space of solutions to the field equations modulo gauge symmetries. This space is now described as (part of) the cohomology ring of a differential complex known as the {\em BV complex},
    \begin{equation}
        \cdots\ \xrightarrow{~Q_{\rm BV}}\ \CCC^\infty_{-1}(\frF)\ \xrightarrow{~Q_{\rm BV}}\ \CCC^\infty_{0}(\frF)\ \xrightarrow{~Q_{\rm BV}}\ \CCC^\infty_{1}(\frF)\ \xrightarrow{~Q_{\rm BV}}\ \cdots~,
    \end{equation}
    where $\CCC^\infty_i(\frF)$ denote functionals of degree~$i$ on the $\RZ$-graded vector space of BV fields $\frF$, which is parametrised by fields, ghosts (of positive degree), and anti-fields (of negative degree). Since the image of $Q_{\rm BV}$ on antifields produces the equations of motion, 
    \begin{equation}
        Q_{\rm BV} \phi_I^+\ =\ \frac{\delta S[\phi^I]}{\delta \phi^I}\ \stackrel{!}{=}\ 0
    \end{equation}
    for $\phi^I$ the classical fields, $\phi_I^+$ their antifields and $S[\phi^I]$ the classical action, the image of the operator $Q_{\rm BV}$ restricted to elements of $\CCC^\infty_{-1}(\frF)$ linear in the antifields is the ideal of functionals in $\CCC^\infty_0(\frF)$ vanishing on solutions of the field equations. Gauge invariant such functionals are in the kernel of the BV differential $Q_{\rm BV}:\CCC^\infty_0(\frF)\rightarrow \CCC^\infty_1(\frF)$. The elements of degree~0 in the cohomology ring thus contain indeed the classical observables.
    
    \paragraph{$L_\infty$-algebra structure.}
    Now the BV complex is evidently a differential graded commutative algebra and its dual is a differential graded commutative coalgebra, which is nothing but an $L_\infty$-algebra. Explicitly, the action of the BV differential $Q_{\rm BV}$ on the coordinate functions on $\frF$ is written as a polynomial in the fields, ghosts, antighosts, and their derivatives, which is simply the dual of the sum over all higher products $\mu_i$ on the graded vector space $\frF$. Schematically, we can write
    \begin{equation}\label{eq:Q_mu}
        Q_{\rm BV} \xi\ =\ -\sum_{i\geq 1}\frac{1}{i!}\mu_i(\xi,\dots,\xi)~,
    \end{equation}
    where $\xi$ is the sum over all coordinate functions on $\frF$ parametrising fields, ghosts, and antighosts. Given the $\mu_i$, we can construct $Q_{\rm BV}$ and knowing $Q_{\rm BV}$, we can reconstruct the $\mu_i$ as higher products from this equation.
    
    Recall furthermore that the BV formalism comes with a Poisson bracket, sometimes called the {\em antibracket}, which is induced by a canonical symplectic form of degree~$-1$ on $\frF$. This form induces a cyclic structure on the $L_\infty$-algebra on $\frF$, which is of degree~$-3$.
    
    Altogether, we conclude that any field theory has an associated cyclic $L_\infty$-algebra structure on its BV field space $\frF$, and this structure is recovered from the BV differential and the BV antibracket. See~\cite{Jurco:2018sby,Jurco:2019bvp}~for more details and the precise formulation of equation~\eqref{eq:Q_mu} and for a review about the $Q$-manifold formulation of $L_\infty$-algebras.
    
    The fact that any classical Lagrangian field theory comes with an $L_\infty$-algebra is certainly well-known by experts on the BV formalism. It has been rediscovered several times, see e.g.~\cite{IuliuLazaroiu:2009wz} or~\cite{Hohm:2017pnh} and also~\cite{Jurco:2018sby,Jurco:2019bvp} for more historical references. The structural advantages of this description, however, have not been fully exploited in our opinion, and this is what we set out to do in this paper.
    
    \section{Scalar field theory}\label{sec:scalar_FT}
    
    As an introductory example illustrating the construction of an $L_\infty$-algebra for a classical field theory, the computation of its minimal model, and the recursion relations, we consider scalar field theory on four-dimensional Minkowski space $\MM\coloneqq(\FR^4,\eta)$ with $\eta$ the Minkowski metric. In the following, $\mu,\nu,\ldots=0,\ldots,3$, and we shall write $x\cdot y\coloneqq\eta_{\mu\nu}x^\mu y^\nu=x_\mu y^\mu$ and $\Box\coloneqq\partial^\mu\partial_\mu$. All of our constructions in this section generalise rather evidently to arbitrary field theories admitting a (classical) BV formulation.
    
    \subsection{\texorpdfstring{$L_\infty$}{Linfinity}-algebra formulation of scalar field theory}
    
    Instead of plain $\varphi^4$-theory, we start from the action 
    \begin{equation}\label{eq:phi4-action}
        S\ \coloneqq\ \int \dd^{4} x~\left\{\tfrac12 \varphi(\Box-m^2) \varphi-\tfrac{\kappa}{3!}\varphi^3-\tfrac{\lambda}{4!}\varphi^4\right\},
    \end{equation}
    which will demonstrate the relation between the minimal model and tree level amplitudes more clearly.
    
    \paragraph{\mathversion{bold}Scalar $L_\infty$-algebra.}
    The associated $L_\infty$-algebra of this field theory is obtained as usual from the BV formalism.\footnote{See also~\cite{Zeitlin:2007fp} for pure $\varphi^4$-theory and~\cite{Jurco:2018sby} for a discussion closer to ours.} Here, we merely note that in a field theory without (gauge) symmetry to be factored out, the BV action agrees with the classical action. The homological vector field $Q_{\rm BV}$ therefore acts only non-trivially on the anti-field $\varphi^+$, and we have
    \begin{equation}
        Q_{\rm BV} \varphi^+\ \coloneqq\ \{S_{\rm BV},\varphi^+\}\ =\ \frac{\delta S}{\delta \varphi}\ =\ \sum_{i\geq 1}\frac{1}{i!}\mu_i(\varphi,\dots,\varphi)~.
    \end{equation}
    The resulting $L_\infty$-algebra is therefore
    \begin{subequations}\label{eq:phi4-naive-L_infty}
        \begin{equation}\label{eq:phi4-complex}
            \underbrace{*}_{=:\,\sL_0}\ \xrightarrow{\phantom{-\Box-m^2}}\ \underbrace{\CCC^\infty(\MM)}_{=:\,\sL_1} \ \xrightarrow{~\Box-m^2~}\  \underbrace{\CCC^\infty(\MM)}_{=:\, \sL_2} \ \xrightarrow{\phantom{-\Box-m^2}}\  \underbrace{*}_{=:\,\sL_3}
        \end{equation}
        with higher products
        \begin{equation}
            \begin{gathered}
                \mu_1(\varphi_1)\ \coloneqq\ (\Box-m^2)\varphi_1~,~~
                \mu_2(\varphi_1,\varphi_2)\ \coloneqq\ -\kappa\varphi_1\varphi_2~,\\
                \mu_3(\varphi_1,\varphi_2,\varphi_3)\ \coloneqq\ -\lambda\varphi_1\varphi_2\varphi_3
            \end{gathered}
        \end{equation}
    \end{subequations}
    for $\varphi_{1,2,3}\in \CCC^\infty(\MM)$. The homotopy Maurer--Cartan action~\eqref{eq:hMC_action} for this $L_\infty$-algebra becomes $S$.

    \paragraph{Cyclic structure.}
    This $L_\infty$-algebra, however, is too general. In particular, we cannot extend it to a cyclic one with the cyclic structure given by the integral,
    \begin{equation}
        \langle \varphi_1,\varphi_2\rangle_\sfL\ =\ \int \dd^4 x~\varphi_1(x) \varphi_2(x)~.
    \end{equation}
    Firstly, finiteness of the integral is not guaranteed and secondly, boundary terms arising when partially integrating the Laplace operator may violate cyclicity. We are thus led to restricting the function space to the Schwartz functions $\CCS(\MM)\subseteq \CCC^\infty(\FR^{1,3})$, i.e.~functions which are rapidly decreasing towards the boundary of Minkowski space $\MM$. This restriction, however, is too harsh: the $\eps$-deformed kinematical operator,
    \begin{equation}\label{eq:eps-reg-mu_1}
        \mu_1(\varphi_1)\ \coloneqq\ (\Box-m^2+\di \eps)\varphi_1
    \end{equation} 
    for $\eps\in\FR^+$, is invertible as a map $\mu_1:\CCS(\MM)\rightarrow\CCS(\MM)$, as we will explain in more detail below. Therefore, the cohomology of $\sL$ would be trivial and the $L_\infty$-algebra would be quasi-isomorphic to the trivial one. 
    
    To fix this issue, we should also include the solutions
    to the classical Klein--Gordon equation, $\ker(\mu_1)\subseteq \CCC^\infty(\MM)$, in our field space. More precisely, we should restrict ourselves to those solutions with Schwartz-type Cauchy data, $\ker_\CCS(\mu_1)\subseteq \ker(\mu_1)$. Our total field space is then
    \begin{equation}
        \frF\ \coloneqq\ \ker_\CCS(\mu_1)\oplus \CCS(\MM)~,
    \end{equation}
    where both function spaces are necessarily complexified and at least $\ker_\CCS(\mu_1)$ is extended to distributions. Note that both subspaces are vector spaces and their intersection is empty since there are no solutions to the Klein--Gordon equation in $\CCS(\MM)$ which is a simple consequence of energy conservation. 
    
    This, however, requires an adjustment of inner product and the higher products. The complexificiation of the cyclicity condition~\eqref{eq:cyclicity} is a straightforward extension of the ad-invariance condition from complex quadratic Lie algebras. We can therefore choose the inner product for any $\varphi_{1,2}^{\rm free}\in\ker_\CCS(\mu_1)$ and $\varphi^{\rm int}_{1,2}\in\CCS(\MM)$ to be 
    \begin{subequations}\label{eq:inner_prod_L}
        \begin{equation}
            \langle\varphi^{\rm free}_1+\varphi^{\rm int}_1,\varphi^{\rm free}_2+\varphi^{\rm int}_2\rangle_\sfL\ \coloneqq\ \langle \varphi_1^{\rm int},\varphi_2^{\rm int}\rangle_\CCS+\langle \varphi_1^{\rm free},\varphi_2^{\rm free}\rangle_{\rm ker}
        \end{equation}
        with
        \begin{equation}
            \langle \varphi_1^{\rm int},\varphi_2^{\rm int}\rangle_\CCS\ \coloneqq\ \int \dd^4 x ~(\varphi^{\rm int}_1)^*\varphi^{\rm int}_2
        \end{equation}
        the evident inner product on (the complexification of) $\CCS(\MM)$. On the complexification of $\ker_\CCS(\mu_1)$, a common choice would be the indefinite Klein--Gordon inner product,
        \begin{equation}
            \langle \varphi_1^{\rm free},\varphi_2^{\rm free}\rangle_{\rm ker}\ \coloneqq\ \di\left.\int \dd^3\vec{x} ~\left( (\varphi_1^{\rm free})^*\,\dpar_t \varphi_2^{\rm free}-\varphi_2^{\rm free}\,\dpar_t (\varphi_1^{\rm free})^*\right)\right|_{x=(t,\vec{x})}~.
        \end{equation}
        One may feel uncomfortable that this inner product is indefinite and vanishes for real scalar fields; an alternative would be the positive definite inner product proposed e.g.~in~\cite{Kleefeld:2006bp}. Recall, however, that the Klein--Gordon inner product also features in the proof of the LSZ reduction formula for scalar fields, cf.~e.g.~\cite{Itzykson:1980rh}. 
    \end{subequations}
    
    We also note that $\frF$ is not closed under multiplication: for example, the product between two elements in $\ker_\CCS(\mu_1)$ is neither in $\CCS(\MM)$ nor in $\ker_\CCS(\mu_1)$. To remedy this, we define the maps
    \begin{equation}\label{eq:scalar_higher_prods_a}
        \begin{aligned}
            \mu_2(\varphi_1,\varphi_2)|_{\CCS(\MM)}\ &\coloneqq\ -\kappa\,\de^{-\delta t^2}\varphi_1\varphi_2~,\\
            \mu_3(\varphi_1,\varphi_2,\varphi_3)|_{\CCS(\FR^{1,3})}\ &\coloneqq\ -\lambda\,\de^{-\delta t^2}\varphi_1\varphi_2\varphi_3~.
        \end{aligned}
    \end{equation}
    Clearly, these products break Lorentz invariance, but we can eventually consider the limit $\delta \rightarrow +0$ to restore Lorentz invariance in all our final results. The values these maps take in ${\rm ker}_\CCS(\mu_1)$ can then be determined by cyclicity if at least one of the arguments is an interacting field $\varphi_1^{\rm int}\in \CCS(\MM)$. For example,
    \begin{equation}
        \langle \varphi_2^{\rm free},\mu_2(\varphi_3,\varphi_1^{\rm int})\rangle_{\rm ker}\ \coloneqq\ \langle \varphi_1^{\rm int},\mu_2(\varphi_2^{\rm free},\varphi_3)\rangle_\CCS 
    \end{equation}
    fully fixes the map $\mu_2:\frF\times \CCS(\MM)\rightarrow {\rm ker}_\CCS(\mu_1)$. The remaining components are simply defined via momentum conservation:
    \begin{equation}
        \begin{aligned}
            \mu_2(\varphi_1,\varphi_2)|_{\ker_\CCS(\mu_1)}\ &\coloneqq\ -\kappa\int \frac{\dd^3 \vec{k}}{(2\pi)^3} \frac{1}{2 E_{\vec{k}}}~\de^{\di (E_{\vec{k}} t-\vec{k}\cdot\vec{x})}\int \dd^4 x' ~\de^{-\di (E_{\vec{k}} t'-\vec{k}\cdot\vec{x'})}\varphi_1(x')\varphi_2(x')~,\\
            \mu_3(\varphi_1,\varphi_2,\varphi_3)|_{\ker_\CCS(\mu_1)}\ &\coloneqq\ -\lambda\int \frac{\dd^3 \vec{k}}{(2\pi)^3} \frac{1}{2 E_{\vec{k}}}~\de^{\di (E_{\vec{k}} t-\vec{k}\cdot\vec{x})}~\times\\
            &\hspace{3cm}\times\int \dd^4 x'~\de^{-\di (E_{\vec{k}} t'-\vec{k}\cdot\vec{x'})}\varphi_1(x')\varphi_2(x')\varphi_3(x')~,
        \end{aligned}
    \end{equation}
    for $\varphi_{1,2,3}\in \ker_\CCS(\mu_1)$, where $E_{\vec{k}}\coloneqq\sqrt{|\vec{k}|^2+m^2-\di \eps}$.

    We note that the homotopy Jacobi identities are trivially satisfied since the only non-trivial higher products $\mu_i$ map $\sL_1\times\cdots\times\sfL_1$ to $\sL_2$ and thus nested expressions of $\mu_i$ vanish trivially. Altogether, we defined a cyclic $L_\infty$-algebra
    \begin{equation}\label{eq:scalar-complex}
        \underbrace{*}_{=:\,\sL_0}\ \xrightarrow{\phantom{-\Box-m^2}}\ \underbrace{\frF}_{=:\,\sL_1} \ \xrightarrow{~\Box-m^2+\di\eps~}\  \underbrace{\frF}_{=:\, \sL_2} \ \xrightarrow{\phantom{-\Box-m^2}}\  \underbrace{*}_{=:\,\sL_3}~.
    \end{equation}
    
    It should be rather obvious that the above construction of a cyclic $L_\infty$-algebra can be performed for any field theory to which we can apply the BV formalism; in particular, the specialisation of the field space to Schwartz functions and on-shell modes readily generalises.  We also note that the technicalities of choosing the appropriate field space mostly arose because we insisted on a consistent cyclic structure on the $L_\infty$-algebra. If one is happy to do without a precise cyclic structure, one can essentially neglect this issue.
    
    \subsection{Strictification of scalar field theory}\label{ssec:scalar_strict}
    
    Let us briefly discuss the strictification of the $L_\infty$-algebra $\sL$, which consists of an $L_\infty$-algebra $\tilde \sL$ which is quasi-isomorphic to $\sL$ and for which $\tilde \mu_i=0$ for $i\geq 3$. Equivalently, we find an action $\tilde S$ which yields a field theory that is classically equivalent to that of the action~\eqref{eq:phi4-action} but whose interaction terms are at most cubic in the fields. This is done by introducing auxiliary fields, and the strictification for pure $\varphi^4$-theory with $\kappa=0$ was already given in~\cite{Jurco:2018sby}. For simplicity, we will work with the naive, unregularised $L_\infty$-algebra~\eqref{eq:phi4-naive-L_infty}. 
    
    \paragraph{Differential graded Lie algebra structure.}
    Constructing an equivalent action $\tilde S$ is straightforward, and one possible form is 
    \begin{equation}
        \tilde S\ \coloneqq\ \int \dd^{4} x~\left\{\tfrac12 \varphi(\Box -m^2) \varphi+XY-\tfrac{\kappa}{3!}Y\varphi+\tfrac{4\kappa}{\lambda}X\varphi-\tfrac{\lambda}{4!}Y^2+X\varphi^2\right\},
    \end{equation}
    where $X$ and $Y$ are two additional auxiliary scalar fields $X,Y\in \CCC^\infty(\FR)$. The corresponding $L_\infty$-algebra $\tilde \sL$ reads as
    \begin{subequations}
        \begin{equation}
            \underbrace{*}_{=:\,\tilde \sL_0}\ \xrightarrow{\phantom{-\dpar^\mu\dpar_\mu-m^2}}\ \underbrace{\CCC^\infty(\MM)\otimes \FR^3}_{=:\,\tilde \sL_1} \ \xrightarrow{~~~~\tilde \mu_1~~~~}\  \underbrace{\CCC^\infty(\MM)\otimes \FR^3}_{=:\, \tilde \sL_2} \ \xrightarrow{\phantom{-\dpar^\mu\dpar_\mu-m^2}}\  \underbrace{*}_{=:\,\tilde \sL_3}~,
        \end{equation}
        and has non-trivial higher products
        \begin{equation}
            \begin{gathered}
                \tilde \mu_1(\varphi_1,X_1,Y_1)\ \coloneqq\ \big((\Box-m^2)\varphi_1-\tfrac{\kappa}{3!}Y_1+\tfrac{4\kappa}{\lambda}X_1,Y_1+\tfrac{4\kappa}{\lambda}\varphi,X_1-\tfrac{\kappa}{3!}\varphi_1-\tfrac{\lambda}{12}Y_1\big)~,\\
                \tilde \mu_2\big((\varphi_1,X_1,Y_1),(\varphi_2,X_2,Y_2)\big)\ \coloneqq\ (2\varphi_1 X_2+2\varphi_2 X_1,2\varphi_1\varphi_2,0)~.
            \end{gathered}
        \end{equation}
    \end{subequations}
    for $\varphi,\varphi_i,X,X_i,Y,Y_i\in \CCC^\infty(\MM)$. The homotopy Maurer--Cartan action~\eqref{eq:hMC_action} for $\tilde\sfL$ is indeed $\tilde S$.
    
    A possible quasi-isomorphism $\varphi:\tilde \sL\rightarrow \sL$ has non-trivial maps
    \begin{equation}
        \begin{aligned}
            \phi_1&:\tilde \sL_{1}\ \rightarrow\ \sL_{1}~,~~~&\phi_1(\varphi,X,Y)\ &\coloneqq\ \varphi~,\\
            \phi_1&:\tilde \sL_{2}\ \rightarrow\ \sL_{2}~,~~~&\phi_1(\varphi,X,Y)\ &\coloneqq\ \varphi-\tfrac{\kappa}{3!}X-\tfrac{4\kappa}{\lambda}Y~,\\
            \phi_2&:\tilde \sL_1\times \tilde \sL_2\ \rightarrow\ \sL_2~,~~~&\phi_2\big((\varphi_1,X_1,Y_1),(\varphi_2,X_2,Y_2)\big)\ &\coloneqq\ -2\varphi_1Y_2+\tfrac{\lambda}{3!}\varphi_1 X_2~.
        \end{aligned}
    \end{equation}
    As one readily checks, these maps satisfy the non-trivial relations in~\eqref{eq:L_infty_morphism},
    \begin{equation}
        \begin{aligned}
            i\ =\ 1~&:~\mu_1(\phi_1(\Phi_1))\ =\ \phi_1(\tilde \mu_1(\Phi_1))~,\\
            i\ =\ 2~&:~\mu_1(\phi_2(\Phi_1,\Phi_2))+\mu_2(\phi_1(\Phi_1),\phi_1(\Phi_2))\ =\\
            &\kern1cm =\ \phi_1(\tilde\mu_2(\Phi_1,\Phi_2))\, -\phi_2(\tilde\mu_1(\Phi_1),\Phi_2)-\phi_2(\tilde\mu_1(\Phi_2),\Phi_1)~,\\
            i\ =\ 3~&:~\mu_3(\phi_1(\Phi_1),\phi_1(\Phi_2),\phi_1(\Phi_3))\ = \ -[\phi_2(\tilde \mu_2(\Phi_1,\Phi_2),\Phi_3)+{\rm cyclic}]~,
        \end{aligned}
    \end{equation}
    where $\Phi_i=(\varphi_i,X_i,Y_i)\in \tilde \sL_1$. 
    \subsection{Scattering amplitudes and recursion relations}
    
    \paragraph{Minimal model.}
    To compute the minimal model $\sL^\circ$ of $\sL$, we need to find a contracting homotopy
    \begin{equation}
        \myxymatrix{\ar@(dl,ul)[]^h \sfL~\ar@<+2pt>@{->>}[rr]^{\kern-20pt \sfp} & & ~\sL^\circ\coloneqq H^\bullet_{\mu_1}(\sfL) \ar@<+2pt>@{^(->}[ll]^{\kern-20pt \sfe}}.
    \end{equation}
    We start by noticing that $\mu_1$ is a map $\mu_1:\frF\rightarrow \CCS(\MM)$ and it is invertible as a map $\mu_1:\CCS(\MM)\rightarrow \CCS(\MM)$. Its inverse is the Feynman propagator $G^F$, which is defined for functions $\varphi\in \CCS(\MM)$ by
    \begin{subequations}\label{eq:green_operator}
        \begin{equation}
            (G^F\varphi)(x)\ \coloneqq\ \int \dd^4 y~G^F(x,y) \varphi(y)~,
        \end{equation}
        where the integral kernel is
        \begin{equation}
            G^F(x,y)\ =\ \int \frac{\dd^4 k}{(2\pi)^4}~\de^{-\di k\cdot (x-y)} \hat G^F(k)\ewith \hat G^F(k)\ =\ \frac{1}{k^2+m^2-\di \eps}~.
        \end{equation}
    \end{subequations}
    It satisfies 
    \begin{subequations}
        \begin{equation}
            \left(\Box-m^2+\di\eps\right)(G^F\varphi)\ =\ \varphi~,
        \end{equation}
        for $\varphi\in \CCS(\MM)$ or, more formally,
        \begin{equation}
            G^F\circ \mu_1|_{\CCS(\MM)}\ =\ \mu_1\circ G^F\ =\ \id_{\CCS(\MM)}~.
        \end{equation}
    \end{subequations}
    For more and precise details, see e.g.~\cite[Chapter 14]{Zeidler:2011bk1}. We trivially extend $G^F$ to a linear function $\tilde G^F$ on all of $\frF$ with $\ker(\tilde G^F)=\ker_\CCS(\mu_1)$. 
    
    Because the invertibility of $\mu_1$ on $\CCS(\MM)$ implies surjectivitiy on $\CCS(\MM)$, it is now clear that the graded vector space of the minimal model $\sL^\circ$ of $\sL$ is 
    \begin{equation}
        \sL^\circ\ =\ H^\bullet_{\mu_1}(\sL)\ \cong\ \Big(~~*\ \xrightarrow{~~~~~}\ \ker_\CCS(\mu_1)\ \xrightarrow{~~0~~}\ \ker_\CCS(\mu_1)\ \xrightarrow{~~~~~}\ *~~\Big)~.
    \end{equation}
    With the help of $\tilde G^F$, we can define the projections
    \begin{equation}
        \begin{aligned}
            &\sfp_1:\sL_1\rightarrow \ker_\CCS(\mu_1)~,~~~\sfp_1\ \coloneqq\ \id-\tilde G^F\circ \mu_1~,\\
            &\sfp_2:\sL_2\rightarrow \ker_\CCS(\mu_1)~,~~~\sfp_2\ \coloneqq\ \id-\mu_1\circ \tilde G^F~.
        \end{aligned}
    \end{equation}
    The embeddings $\sfe_{1,2}:\ker_\CCS(\mu_1)\embd \frF$ are simply the trivial ones. The maps we introduced so far satisfy the relations 
    \begin{equation}
        \sfp\circ \sfe\ =\ \id_{\sL^\circ}~,~~~\mu_1\circ \mu_1\ =\ 0~,~~~\mu_1\circ \sfe\ =\ 0~,~~~\sfp\circ \mu_1\ =\ 0
    \end{equation}
    and we have the following picture:
    \begin{equation}
        \vcenter{\vbox{\xymatrix@C=2.5pc@R=0.4pc{
                    &&&&& \CCS(\MM)\ar@/^15pt/@{->}^{\mu_1}[r] & \CCS(\MM)\ar@/^15pt/@{->}_{G^F}[l] \\
                    {*} \ar@{->}[r] & \sL_1 \ar@{->}[r]\ar@/^8pt/@{->}[ddddd]^{\sfp_1} & \sL_2\ar@/^8pt/@{->}[ddddd]^{\sfp_2} \ar@{->}[r]  & {*}  \ar@{}[r]^(.35){}="a"^(.65){}="b" \ar@{=} "a";"b"&{*} \ar@{}[r]^(0.05){}="a1"^(.68){}="b1"\ar@{->} "a1";"b1" & \oplus & \oplus \ar@{}[r]^(0.30){}="a2"^(.95){}="b2"\ar@{->} "a2";"b2"& {*}\\
                    &&&&&  \ker_\CCS(\mu_1)\ar@/^5pt/@{->}[dddd]^{\sfp_1}\ar@{->}[r]^{0}& \ker_\CCS(\mu_1) \ar@/^5pt/@{->}[dddd]^{\sfp_2}\\
                    \\ \\ \\
                    {*} \ar@{->}[r] & \sL^\circ_1 \ar@/^8pt/@{->}[uuuuu]^{\sfe_1} \ar@{->}[r] & \sL^\circ_2 \ar@/^8pt/@{->}[uuuuu]^{\sfe_2}\ar@{->}[r]  & {*}  \ar@{}[r]^(.35){}="c"^(.65){}="d" \ar@{=} "c";"d"&{*} \ar@{->}[r] & \ker_\CCS(\mu_1)\ar@/^5pt/@{->}[uuuu]^{\sfe_1}\ar@{->}[r]^{0}& \ker_\CCS(\mu_1)\ar@/^5pt/@{->}[uuuu]^{\sfe_2}\ar@{->}[r]  & {*}
                }}}
    \end{equation}
    
    It remains to construct a contracting homotopy $h$, i.e.~a map $h:\sL\rightarrow \sL$ of degree~$-1$ such that
    \begin{equation}
        \id_\sL\ =\ \sfe\circ \sfp+h\circ \mu_1+\mu_1\circ h~
    \end{equation}
    and clearly the extended Feynman propagator $\tilde G^F$ satisfies this relation. We thus put $h$ trivial except for $h_2\coloneqq\tilde G^F$. The improved contracting homotopy $\sfh$, cf.~\eqref{eq:improvement_h}, agrees with $h$, 
    \begin{equation}
        \sfh\ =\ h~,
    \end{equation}
    since $h^2=0$ and $h\circ \sfe=0$. The decomposition of $\sL$ reads as 
    \begin{equation}
        \begin{aligned}
            \sL_1\ &\cong\ \sL_{{\rm harm}, 1}\oplus \sL_{{\rm coex},1}\ \cong\ \ker_\CCS(\mu_1)\oplus \CCS(\MM)~,\\
            \sL_2\ &\cong\ \sL_{{\rm harm},2}\oplus \sL_{{\rm ex},2}\ \cong\ \ker_\CCS(\mu_1)\oplus \CCS(\MM)~.
        \end{aligned}
    \end{equation}
    
    The minimal model is now readily computed using the formulas~\eqref{eq:minimal_model}:
    \begin{equation}
        \begin{aligned}
            \mu^\circ_2(\varphi_1,\varphi_2)\ &=\ (\sfp\circ\mu_2)(\varphi_1,\varphi_2)~,\\
            \mu^\circ_3(\varphi_1,\varphi_2,\varphi_3)\ &=\
            -\sum_{\sigma\in {\rm Sh(1;3)}}(\sfp\circ\mu_2)\big(\sfe(\varphi_{\sigma(1)}),\tilde G^F(\mu_2(\sfe(\varphi_{\sigma(2)}),\sfe(\varphi_{\sigma(3)})))\big)\,+\\
            &\kern2cm+(\sfp\circ\mu_3)\big(\sfe(\varphi_1),\sfe(\varphi_2),\sfe(\varphi_3)\big)~,
        \end{aligned}
    \end{equation}
    where $G^F_{x,x'}$ indicates the operator $G^F$ acting on a function of $y$ and producing a function of $x$. Similarly, $\mu_4$ reads as
    \begin{equation}
        \begin{aligned}
            &\mu^\circ_4(\varphi_1,\varphi_2,\varphi_3,\varphi_4)\ =\\
            &\kern.5cm=\ \sum_{\sigma\in {\rm Sh(1,1;4)}}(\sfp\circ\mu_2)\big(\varphi_{\sigma(1)},\tilde G^F(\mu_2(\varphi_{\sigma(2)},\tilde G^F(\mu_2(\varphi_{\sigma(3)},\varphi_{\sigma(4)}))))\big)\,-\\
            &\kern2cm-\sum_{\sigma\in {\rm Sh(1;4)}}(\sfp\circ\mu_2)\big(\varphi_{\sigma(1)},\tilde G^F(\mu_3(\varphi_{\sigma(2)},\varphi_{\sigma(3)},\varphi_{\sigma(4)})))\big)\,+\\
            &\kern2cm+\frac12\sum_{\sigma\in {\rm Sh(2;4)}}(\sfp\circ\mu_2)\big(\tilde G^F(\mu_2(\varphi_{\sigma(1)},\varphi_{\sigma(2)})),\tilde G^F(\mu_2(\varphi_{\sigma(3)},\varphi_{\sigma(4)})))\big)\,-\\
            &\kern2cm-\sum_{\sigma\in {\rm Sh(2;4)}}(\sfp\circ\mu_3)\big(\varphi_{\sigma(1)},\varphi_{\sigma(2)},\tilde G^F(\mu_2(\varphi_{\sigma(3)},\varphi_{\sigma(4)}))\big)~.
        \end{aligned}
    \end{equation}
    
    \paragraph{Cyclic structure.}
    We also stress that the compatibility of the $L_\infty$-algebra morphism $\phi:\sL^\circ \rightarrow \sL$ with the cyclic structure $\langle -,-\rangle_\sfL$ on $\sL$ is trivially satisfied. First of all, we have $\langle \phi_1(\cdots),\phi_i(\cdots)\rangle_\sfL=\langle \sfe(\cdots),\sfh(\cdots)\rangle_\sfL=0$ for $i\geq 2$ since $\im(\sfe)\cong\ker_\CCS(\MM)$, $\im(\sfh)\cong\CCS(\MM)$ and the direct sum $\frF=\ker_\CCS(\MM)\oplus \CCS(\MM)$ is an orthogonal decomposition with respect to the cyclic structure $\langle -,-\rangle_\sfL$. Then, we have $\langle \phi_i(\cdots),\phi_j(\cdots)\rangle_\sfL=\langle \sfh(\cdots),\sfh(\cdots)\rangle_\sfL=0$ for $i,j\geq 2$, since $\im(\sfh)\subseteq \sL_2$ and the cyclic structure $\langle-,-\rangle_\sfL$ necessarily vanishes between two elements of $\sL_2$.
    
    At a more abstract level and for a general field theory, it is rather evident that we will always have a Feynman propagator serving as a contracting homotopy, which allows us to construct the minimal model of the field theory's $L_\infty$-algebra explicitly. Also, the graded vector space $\sL_1$ will consist of the on-shell modes, $\sL_2$ will be isomorphic to $\sL_1$ and the symplectic form on field space provides a non-degenerate pairing of both spaces.

    \paragraph{Scattering amplitudes.}
    Let us now link the minimal model to tree-level Feynman diagrams. We start from the summands in the homotopy Maurer--Cartan action~\eqref{eq:hMC_action} for the minimal model,
    \begin{equation}
        \frac{1}{i!}\langle \varphi,\mu^\circ_i(\varphi,\dots,\varphi)\rangle_{\sfL^\circ}~,
    \end{equation}
    where the inner product $\langle-,-\rangle_{\sfL^\circ}$ is given by $\langle -,-\rangle_{{\rm ker}_\CCS(\mu_1)}$ defined in~\eqref{eq:inner_prod_L}. Using\linebreak the usual trick of polarisation, we can introduce the $(i+1)$-point functions\linebreak $\langle \varphi_1,\mu_i^\circ(\varphi_2,\dots,\varphi_{i+1})\rangle_{\sfL^\circ}$.
    As we shall see now, these are indeed the tree-level $(i+1)$-point functions. Let us consider the 3-point function $\frac{1}{3!}\langle \varphi_1,\mu^\circ_2(\varphi_2,\varphi_3)\rangle^\circ$ for $\varphi_{1,2,3}$ plane waves with on-shell momenta $k_1$, $k_2$, and $k_3$, respectively. We have
    \begin{equation}
        \langle \varphi_1,\mu^\circ_2(\varphi_2,\varphi_3)\rangle_{\sfL^\circ}\ =\ -\kappa(2\pi)^4\delta(k_1+k_2+k_3)~,
    \end{equation}
    and for $m>0$, this amplitude vanishes.
    
    For the 4-point function we now get additional contributions from the 3-point vertices:
    \begin{equation}
        \begin{aligned}
            &\langle \varphi_1,\mu^\circ_3(\varphi_2,\varphi_3,\varphi_4)\rangle_{\sfL^\circ}\ =\ \\
            &\kern1cm=\ \langle \varphi_4,\mu^\circ_3(\varphi_1,\varphi_2,\varphi_3)\rangle_{\sfL^\circ}\\
            &\kern1cm=\ \Big\langle \varphi_4~,~ -\sum_{\sigma\in {\rm Sh(1;3)}}(\sfp\circ\mu_2)\big(\sfe(\varphi_{\sigma(1)}),\tilde G^F(\mu_2(\sfe(\varphi_{\sigma(2)}),\sfe(\varphi_{\sigma(3)})))\big)\,+\\
            &\kern3cm+(\sfp\circ\mu_3)\big(\sfe(\varphi_1),\sfe(\varphi_2),\sfe(\varphi_3)\big)\Big\rangle\\
            &\hspace{1cm}=\ -(2\pi)^4\delta(k_4+k_1+k_2+k_3)\left(\kappa^2\sum_{\sigma\in {\rm Sh(1;3)}}\frac{1}{(k_{\sigma(2)}+k_{\sigma(3)})^2+m^2-\di \eps}+\lambda\right)~.
        \end{aligned}
    \end{equation}
    In terms of Feynman diagrams, we have the 4-point function
    \begin{equation}
        \vcenter{\hbox{
                \begin{axopicture}{(80,80)(0,0)}
                    \SetArrowStroke{0.5}
                    \Vertex(55,40){2}
                    \Vertex(25,40){2}
                    \Line(25,40)(55,40)
                    \Line(10,10)(25,40)
                    \Line(10,70)(25,40)
                    \Line(70,10)(55,40) 
                    \Line(70,70)(55,40) 
                    \Text(5,5){$\varphi_1$}
                    \Text(75,5){$\varphi_2$}
                    \Text(5,75){$\varphi_4$}
                    \Text(75,75){$\varphi_3$}
                    \Text(15,40){$\kappa$}
                    \Text(65,40){$\kappa$}
                \end{axopicture}
                ~~~
                \begin{axopicture}{(80,80)(0,0)}
                    \SetArrowStroke{0.5}
                    \Vertex(40,55){2}
                    \Vertex(40,25){2}
                    \Line(40,25)(40,55)
                    \Line(10,10)(40,25)
                    \Line(10,70)(40,55)
                    \Line(70,10)(40,55) 
                    \Line(70,70)(40,25) 
                    \Text(5,5){$\varphi_1$}
                    \Text(75,5){$\varphi_2$}
                    \Text(5,75){$\varphi_4$}
                    \Text(75,75){$\varphi_3$}
                    \Text(40,15){$\kappa$}
                    \Text(40,65){$\kappa$}
                \end{axopicture}
                ~~~
                \begin{axopicture}{(80,80)(0,0)}
                    \SetArrowStroke{0.5}
                    \Vertex(40,55){2}
                    \Vertex(40,25){2}
                    \Line(40,25)(40,55)
                    \Line(10,10)(40,25)
                    \Line(10,70)(40,55)
                    \Line(70,10)(40,25) 
                    \Line(70,70)(40,55) 
                    \Text(5,5){$\varphi_1$}
                    \Text(75,5){$\varphi_2$}
                    \Text(5,75){$\varphi_4$}
                    \Text(75,75){$\varphi_3$}
                    \Text(40,15){$\kappa$}
                    \Text(40,65){$\kappa$}
                \end{axopicture}
                ~~~
                \begin{axopicture}{(80,80)(0,0)}
                    \SetArrowStroke{0.5}
                    \Vertex(40,40){2}
                    \Line(10,10)(40,40)
                    \Line(10,70)(40,40)
                    \Line(70,10)(40,40) 
                    \Line(70,70)(40,40) 
                    \Text(5,5){$\varphi_1$}
                    \Text(75,5){$\varphi_2$}
                    \Text(5,75){$\varphi_4$}
                    \Text(75,75){$\varphi_3$}
                    \Text(40,50){$\lambda$}
                \end{axopicture}
            }}
    \end{equation}
    
    Let us now discuss the general interpretation of the $\phi_i$ and the $\mu_i^\circ$ arising in the quasi-isomorphism. While $\phi_1$ is simply the embedding of on-shell modes into the original $L_\infty$-algebra $\sL$, the higher $\phi_i$ take on-shell (i.e.~elements of $\ker_\CCS(\mu_1)$) or off-shell modes (i.e.~elements of $\CCS(\MM)$), combine them with a $(j+1)$-point vertex encoded in $\mu_j$ and propagate the resulting state to an off-shell mode or current in $\CCS(\MM)$. For example,
    \begin{equation}
        \phi_2(\varphi_1,\varphi_2)\ =\ \frac{1}{2}\vcenter{\hbox{\begin{axopicture}{(80,80)(0,0)}
                    \SetArrowStroke{0.5}
                    \Vertex(40,32){2}
                    \Line(40,32)(40,70)
                    \Line(10,10)(40,32)
                    \Line(70,10)(40,32) 
                    \Text(5,5){$\varphi_1$}
                    \Text(75,5){$\varphi_2$}
                    \Text(40,22){$\mu_2$}
                    \Text(50,60){$\tilde G^F$}
                \end{axopicture}}}+\frac{1}{2}\vcenter{\hbox{\begin{axopicture}{(80,80)(0,0)}
                    \SetArrowStroke{0.5}
                    \Vertex(40,32){2}
                    \Line(40,32)(40,70)
                    \Line(10,10)(40,32)
                    \Line(70,10)(40,32) 
                    \Text(5,5){$\varphi_2$}
                    \Text(75,5){$\varphi_1$}
                    \Text(40,22){$\mu_2$}
                    \Text(50,60){$\tilde G^F$}
                \end{axopicture}}}
    \end{equation}
    
    The $\mu_i^\circ$, on the other hand, take either on-shell states or the currents arising from the $\phi_j$ with $j\geq 2$, combine them with a $(j+1)$-point vertex encoded in $\mu_j$ and project the result back to an on-shell state. For example,
    \begin{equation}
        \mu^\circ_2(\varphi_1,\varphi_2)\ =\ \frac{1}{2}\vcenter{\hbox{\begin{axopicture}{(80,80)(0,0)}
                    \SetArrowStroke{0.5}
                    \Vertex(40,32){2}
                    \Line(40,32)(40,70)
                    \Line(10,10)(40,32)
                    \Line(70,10)(40,32) 
                    \Text(5,5){$\varphi_1$}
                    \Text(75,5){$\varphi_2$}
                    \Text(40,22){$\mu_2$}
                    \Text(47,60){$\sfp$}
                \end{axopicture}}}+\frac{1}{2}\vcenter{\hbox{\begin{axopicture}{(80,80)(0,0)}
                    \SetArrowStroke{0.5}
                    \Vertex(40,32){2}
                    \Line(40,32)(40,70)
                    \Line(10,10)(40,32)
                    \Line(70,10)(40,32) 
                    \Text(5,5){$\varphi_2$}
                    \Text(75,5){$\varphi_1$}
                    \Text(40,22){$\mu_2$}
                    \Text(47,60){$\sfp$}
                \end{axopicture}}}
    \end{equation}
    
    \paragraph{Scattering amplitude recursion relations.}
    The tree level $(i+1)$-point functions are now obtained from the inner product of one external state with $\mu^\circ_i$ of the remaining $i$ external states, so their information is encoded in the higher products of the minimal model $\sL^\circ$. The tree-level corrections to the classical $n+1$-point vertex are now evidently constructed from diagrams involving the currents $\phi_j$ with $j<n$, which are recursively constructed from  currents $\phi_k$ with $k<j$. This is obvious from formulas~\eqref{eq:minimal_model} and in terms of Feynman diagrams, we have for example
    \begin{equation}
        \phi_3(\varphi_1,\varphi_2,\varphi_3)\ =\ \vcenter{\hbox{\begin{axopicture}{(80,80)(0,0)}
                    \SetArrowStroke{0.5}
                    \Vertex(40,32){2}
                    \Line(40,32)(40,70)
                    \Line(10,10)(40,32)
                    \Line(40,10)(40,32)
                    \Line(70,10)(40,32) 
                    \Text(5,5){$\varphi_1$}
                    \Text(40,5){$\varphi_2$}
                    \Text(75,5){$\varphi_3$}
                    \Text(50,38){$\mu_3$}
                    \Text(50,60){$\tilde G^F$}
                \end{axopicture}}}~~~+~\frac{1}{2!}\vcenter{\hbox{\begin{axopicture}{(80,80)(0,0)}
                    \SetArrowStroke{0.5}
                    \Vertex(40,32){2}
                    \Line(40,32)(40,70)
                    \Line(10,10)(40,32)
                    \Line(70,10)(40,32) 
                    \Text(5,5){$\phi_2(\varphi_1,\varphi_2)$}
                    \Text(75,5){$\varphi_1$}
                    \Text(50,38){$\mu_2$}
                    \Text(50,60){$\tilde G^F$}
                \end{axopicture}}}
        ~~~+~\cdots
    \end{equation}
    In most interesting quantum field theories, the non-trivial higher products are very restricted, and the recursive computation of the quasi-isomorphism to the minimal model~\eqref{eq:quasi_iso_phi} simplifies to interesting recursion relations for the currents $\phi_i$. These, in turn, may be solved in particular examples, yielding vast simplifications in the evaluation of the tree-level $(i+1)$-point functions. We shall discuss an important example in the next section but, again, it should be clear that our discussion applies to an arbitrary BV quantisable field theory.
    
    \section{Yang--Mills theory}\label{sec:YM}
    
    \subsection{\texorpdfstring{$L_\infty$}{Linfinity}-algebra formulation of Yang--Mills theory}
    
    We now study $\mathfrak{su}(N)$ Yang--Mills theory on four-dimensional Minkowski space $\FR^{1,3}$. Let $\Omega^p(\FR^{1,3},\mathfrak{su}(N))$ be the $\mathfrak{su}(N)$-valued differential $p$-forms on $\FR^{1,3}$. Furthermore, we let $\dd$ be the exterior derivative and set $\dd^\dagger\coloneqq{\star\dd\star}$ for $\star$ the Hodge star operator with respect to the Minkowski metric. To keep our discussion clear, we shall neglect the intricacies of fall-off conditions on our function spaces; the details developed in the example of scalar field theory can be translated to Yang--Mills theory.
    
    \paragraph{\mathversion{bold}Yang--Mills $L_\infty$-algebra.}
    Following~\cite{Movshev:2003ib,Movshev:2004aw,Zeitlin:2007vv,Zeitlin:2007yf}, we consider the $L_\infty$-algebra $(\sL_{\rm YM_2},\mu_i)$ that is defined by the chain complex
    \begin{subequations}\label{eq:ym2_L_algebra}
        \begin{equation}\label{eq:ym2_complex}
            \begin{aligned}
                & \underbrace{\Omega^0(\FR^{1,3},\mathfrak{su}(N))}_{=:\,\sL_0}\ \xrightarrow{~\mu_1\,\coloneqq\,\dd~}\  \underbrace{\Omega^1(\FR^{1,3},\mathfrak{su}(N))}_{=:\,\sL_1}\\
                &\kern2cm  \xrightarrow{~\mu_1\,\coloneqq\,\dd^\dagger\dd~}\  \underbrace{\Omega^1(\FR^{1,3},\mathfrak{su}(N))}_{=:\,\sL_2}\ \xrightarrow{~\mu_1\,\coloneqq\,\dd^\dagger~}\  \underbrace{\Omega^0(\FR^{1,3},\mathfrak{su}(N))}_{=:\,\sL_3}~,
            \end{aligned}
        \end{equation}
        together with the non-vanishing higher products
        \begin{equation}\label{eq:ym2_brackets}
            \begin{gathered}
                \mu_1(c_1)\ \coloneqq\ \dd c_1~,~~~
                \mu_1(A_1)\ \coloneqq\ \dd^\dagger\dd A_1~,~~~
                \mu_1(A^+_1)\ \coloneqq\ \dd^\dagger A^+_1~,\\
                \mu_2(c_1,c_2)\ \coloneqq\ [c_1,c_2]~,~~~ 
                \mu_2(c_1,A_1)\ \coloneqq\ [c_1,A_1]~,\\ 
                \mu_2(c_1,A^+_2)\ \coloneqq\ [c_1,A^+_2]~,~~~
                \mu_2(c_1,c^+_2)\ \coloneqq\ [c_1,c^+_2]~,\\ 
                \mu_2(A_1,A^+_2)\ \coloneqq\ [A_1,A^+_2]~,\\ 
                \mu_2(A_1,A_2)\ \coloneqq\ \dd^\dagger[A_1,A_2]+{\star[A_1,{\star\dd A_2}]}+{\star[A_2,{\star\dd A_1}]}~,\\
                \mu_3(A_1,A_2,A_3)\ \coloneqq\ {\star[A_1,\star[A_2,A_3]]}+{\star[A_2,\star[A_3,A_1]]}+{\star[A_3,\star[A_1,A_2]]}~,
            \end{gathered}
        \end{equation}
    \end{subequations}
    where $[-,-]$ denotes the Lie bracket on $\mathfrak{su}(N)$ with the wedge product understood. Furthermore, $c_{1,2}\in \sfL_0$, $A_{1,2,3}\in\sfL_1$, $A_{1,2}^+\in\sfL_2$, and $c_{2}^+\in \sfL_3$, respectively.
    
    Importantly, the $L_\infty$-algebra $(\sL_{\rm YM_2},\mu_i)$ carries a natural cyclic structure which is non-trivial only for $|\omega_1|_{\sL_{\rm YM_2}}+|\omega_2|_{\sL_{\rm YM_2}}=3$ and then reads as
    \begin{equation}\label{eq:inner_product_YM2}
        \langle \omega_1,\omega_2\rangle_{\sL_{\rm YM_2}}\ \coloneqq\ \int\tr(\omega_1\wedge{\star\omega_2})~.
    \end{equation}
    
    This cyclic $L_\infty$-algebra arises from the classical part of the BV-formalism as explained in Section~\ref{ssec:FT_from_L_infty}, cf.~also~\cite{Jurco:2018sby} for all details. In particular, the relation between the higher products $\mu_i$ and the BV-differential $Q_{\rm BV}$ is given in formula~\eqref{eq:Q_mu}.
    
    \paragraph{Yang--Mills action and Yang--Mills equation.}
    For $a\in\sL$ of degree~$1$ we have $a=A\in\Omega^1(\FR^{1,3},\frg)$ and thus
    \begin{equation}
        \begin{aligned}
            \tfrac12\langle a,\mu_1(a)\rangle_{\sL_{\rm YM_2}}\ &=\ \tfrac12\int \tr(\dd A\wedge {\star\dd A})~,\\
            \tfrac{1}{3!}\langle a,\mu_2(a,a)\rangle_{\sL_{\rm YM_2}}\ &=\ \tfrac12\int \tr(\dd A\wedge {\star[A,A]})\\
            & =\ \tfrac14\int\tr([A,A]\wedge{\star\dd A}+ \dd A\wedge {\star[A,A]})~,\\
            \tfrac{1}{4!}\langle a,\mu_3(a,a,a)\rangle_{\sL_{\rm YM_2}}\ &=\ \tfrac18\int \tr( [A,A]\wedge {\star[A,A]})~.
        \end{aligned}
    \end{equation}
    Consequently, the homotopy Maurer--Cartan action~\eqref{eq:hMC_action} becomes the Yang--Mills action,
    \begin{equation}\label{eq:second_order_ym_action}
        S_{\rm MC}\ =\ \tfrac12\int\tr(F\wedge{\star F})\ewith F\ \coloneqq\ \dd A+\tfrac12[A,A]~.
    \end{equation}
    Furthermore, the curvature~\eqref{eq:curvature} reads as
    \begin{equation}
        f\ =\ {\star\nabla{\star F}}\ \in\ \Omega^1(\FR^{1,3},\mathfrak{su}(N))~,
    \end{equation}
    where $\nabla{\star F}\coloneqq\dd{\star F}+[A,{\star F}]$ so that $f=0$ is, in fact, the Yang--Mills equation. The gauge transformations~\eqref{eq:gauge_trafo} reduce to the standard ones,
    \begin{equation}
        \delta_cA\ =\ \nabla c\eand
        \delta_c F\ =\ -[c,F]
    \end{equation}
    for $c\in\Omega^0(\FR^{1,3},\mathfrak{su}(N))$.
    
    \subsection{Strictification of Yang--Mills theory}\label{ssec:YM_strict}
    
    \paragraph{First-order formulation.}
    It is well-known that four-dimensional Yang--Mills theory admits an alternative \emph{first-order} formulation~\cite{Okubo:1979gt} given by the action functional 
    \begin{equation}\label{eq:first_order_ym_action}
        S\ \coloneqq\ \int\tr(F\wedge B_++\tfrac\eps2 B_+\wedge B_+)~.
    \end{equation}
    Here, $B_+\in\Omega^2_+(\FR^4,\mathfrak{su}(N))$ is an $\mathfrak{su}(N)$-valued self-dual 2-form on $\FR^4$ for $\eps\in\FR^+$ and we switched to Euclidean space to allow for real self-dual 2-forms. As we are only concerned with scattering amplitudes, which depend holomorphically on the kinematic variables, this switch in signature is largely irrelevant.
    
    Integrating out $B_+$, we find
    \begin{equation}
        S\ =\ -\tfrac{1}{2\eps}\int\tr(F_+\wedge F_+)\ =\ -\tfrac{1}{4\eps}\int\tr(F\wedge{\star F})-\tfrac{1}{4\eps}\int\tr(F\wedge F)~,
    \end{equation}
    where $F_+\coloneqq\frac12(F+{\star F})$. Hence, we recover the standard Yang--Mills action~\eqref{eq:second_order_ym_action} plus a topological term, which is irrelevant for perturbation theory. 
    
    \paragraph{\mathversion{bold}Differential graded Lie algebra structure.}
    Note that~\eqref{eq:first_order_ym_action} is only cubic in the interactions and hence the corresponding equations of motion are at most quadratic. The $L_\infty$-algebra $(\sfL_{\rm YM_1},\mu_i)$ corresponding to this action should thus be strict. Indeed, we find the complex (cf.~\cite{Costello:2007ei})
    \begin{subequations}\label{eq:ym1_L_algebra}
        \begin{equation}\label{eq:ym1_complex}
            \begin{aligned}
                &\underbrace{\Omega^0(\FR^4,\mathfrak{su}(N))}_{=:\,\sL_0}\ \xrightarrow{~\mu_1\,\coloneqq\,\dd~}\  \underbrace{\Omega^2_+(\FR^4,\mathfrak{su}(N))\oplus\Omega^1(\FR^4,\mathfrak{su}(N))}_{=:\,\sL_1}\\
                &\kern1cm\xrightarrow{~\mu_1\,\coloneqq\,(\eps+\dd)+P_+\dd~}\ \underbrace{\Omega^2_+(\FR^4,\mathfrak{su}(N))\oplus\Omega^3(\FR^4,\mathfrak{su}(N))}_{=:\,\sL_2}\ \xrightarrow{~\mu_1\,\coloneqq\,0+\dd~}\  \underbrace{\Omega^4(\FR^4,\mathfrak{su}(N))}_{=:\,\sL_3}
            \end{aligned}
        \end{equation}
        together with the higher products~\cite{Rocek:2017xsj,Jurco:2018sby,Jurco:2019bvp}
        \begin{equation}
            \begin{gathered}
                \mu_1(c_1)\ \coloneqq\ \dd c_1~,~~~
                \mu_1(B_{+1}+A_1)\ \coloneqq\ (\eps B_{+1}+P_+\dd A_1)+\dd B_{+1}~,~~~
                \mu_1(A^+_1)\ \coloneqq\ \dd A^+_1~,\\
                \mu_2(c_1,c_2)\ \coloneqq\ [c_1,c_2]~,~~~ 
                \mu_2(c_1,B_{+1}+A_1)\ \coloneqq\ [c_1,B_{+1}]+[c,A_1]~,\\ 
                \mu_2(c_1,B_{+1}^++A^+_1)\ \coloneqq\ [c_1,B_{+1}^+]+[c,A^+_1]~,~~~
                \mu_2(c_1,c^+_2)\ \coloneqq\ [c_1,c^+_2]~,\\ 
                \mu_2(B_{+1}+A_1,B_{+2}+A_2)\ \coloneqq\ P_+[A_1,A_2]+[A_1,B_{+2}]+[A_2,B_{+1}]~,\\
                \mu_2(B_{+1}+A_1,B_{+2}^++A^+_2)\ \coloneqq\ [A_1,A_2^+]+[B_1,B_{+2}^+]~.\\ 
            \end{gathered}
        \end{equation}
    \end{subequations}
    Here, $P_+=\frac12(1+\star)$ and $c_i\in\sL_0$, $(B_{+i}+A_i)\in\sL_1$, $(B_{+i}^++A_i^+)\in\sL_2$, and $c_i^+\in\sL_3$ for $i=1,2$. 
    
    Also $(\sfL_{\rm YM_1},\mu_i)$ can be made cyclic by introducing the degree~$-3$ inner product
    \begin{equation}\label{eq:InnerProductYM1}
        \langle \omega_1,\omega_2\rangle_{\sL_{\rm YM_1}}\ \coloneqq\ \int\tr(\omega_1\wedge\omega_2)~.
    \end{equation}
    It is now a straightforward exercise to check that the homotopy Maurer--Cartan\linebreak action~\eqref{eq:hMC_action} for the $L_\infty$-algebra $(\sfL_{\rm YM_1},\mu_i,\langle-,-\rangle_{\sfL_{\rm YM_1}})$ reduces to the first-order Yang--Mills action~\eqref{eq:first_order_ym_action}.
    
    \paragraph{\mathversion{bold}Quasi-isomorphism.} Whilst we have already seen that the actions~\eqref{eq:second_order_ym_action} and~\eqref{eq:first_order_ym_action} are equivalent by integrating out the self-dual 2-form, it is instructive to give the explicit quasi-isomorphism between $(\sfL_{\rm YM_2},\mu_i,\langle-,-\rangle_{\sfL_{\rm YM_2}})$ and $(\sfL_{\rm YM_1},\mu_i,\langle-,-\rangle_{\sfL_{\rm YM_1}})$.\footnote{Here, we consider $(\sfL_{\rm YM_2},\mu_i,\langle-,-\rangle_{\sfL_{\rm YM_2}})$ on $\FR^4$ as well.}  In particular, we have
    \begin{subequations}
        \begin{equation}\label{eq:QuasiIsoYM12Diagram}
            \xymatrixrowsep{0.2pc}
            \myxymatrix{
                &  \Omega^2_+(\FR^4,\mathfrak{su}(N)) \ar@{->}[r]^{\eps}  \ar@/^1.1pc/[ddr]_{\dd} & \Omega^2_+(\FR^4,\mathfrak{su}(N)) &\\
                \Omega^0(\FR^4,\mathfrak{su}(N))\ar@{->}[dr]^{\dd} \ar@{=}[ddddd]^{} & \oplus & \oplus & \Omega^4(\FR^4,\mathfrak{su}(N)) \\
                & \Omega^1(\FR^4,\mathfrak{su}(N)) \ar@/_1.1pc/[uur]^{P_+\dd}&    \Omega^3(\FR^4,\mathfrak{su}(N))\ar@{->}[ur]^{\dd}& \\
                &&&\\
                &&&\\
                &&&\\
                \Omega^0(\FR^4,\mathfrak{su}(N))\ar@{->}[r]^{\dd} \ar@{=}[dddd]^{} & \Omega^1(\FR^4,\mathfrak{su}(N))  \ar@{->}[uuuu]^{\phi_1}  \ar@{->}[r]^-{\dd{\star\dd}} \ar@{=}[dddd]^{} & \Omega^3(\FR^4,\mathfrak{su}(N)) \ar@{->}[uuuu]^{\phi_1}  \ar@{->}[r]^-{\dd} &  \Omega^4(\FR^4,\mathfrak{su}(N))  \ar@{->}[uuuuu]^{\phi_1}\\
                &&&\\
                &&&\\
                &&&\\
                \Omega^0(\FR^4,\mathfrak{su}(N))\ar@{->}[r]^{\dd} &  \Omega^1(\FR^4,\mathfrak{su}(N)) \ar@{->}[r]^{\dd^\dagger\dd}&  \Omega^1(\FR^4,\mathfrak{su}(N))  \ar@{->}[uuuu]^{{\star}}  \ar@{->}[r]^{\dd^\dagger} & \Omega^0(\FR^4,\mathfrak{su}(N)) \ar@{->}[uuuu]^{{-\star}}
            }
        \end{equation}
        where we have combined the two complexes~\eqref{eq:ym2_complex} and~\eqref{eq:ym1_complex}. The maps $\phi_1$ are given by
        \begin{equation}
            \begin{gathered}
                \phi_1\,:\, \Omega^1(\FR^4,\mathfrak{su}(N))\ \to\ \Omega^2_+(\FR^4,\mathfrak{su}(N))\oplus\Omega^1(\FR^4,\mathfrak{su}(N))~,\\
                A\ \mapsto\ -\tfrac{1}{\eps} P_+ \dd A+A~,
            \end{gathered}
        \end{equation}
        together with
        \begin{equation}
            \begin{gathered}
                \phi_1\,:\,\Omega^3(\FR^4,\mathfrak{su}(N))\ \to\ \Omega^2_+(\FR^4,\mathfrak{su}(N))\oplus\Omega^3(\FR^4,\mathfrak{su}(N))~,\\
                C\ \mapsto\ 0-\tfrac{1}{2\eps}C~
            \end{gathered}
        \end{equation}
        and 
        \begin{equation}
            \begin{gathered}
                \phi_1\,:\,\Omega^4(\FR^4,\mathfrak{su}(N))\ \to\ \Omega^4(\FR^4,\mathfrak{su}(N))~,\\
                D\ \mapsto\ -\tfrac{1}{2\eps}D~.
            \end{gathered}
        \end{equation}
    \end{subequations}
    As one may check, all square-subdiagrams of~\eqref{eq:QuasiIsoYM12Diagram} are commutative, and, consequently, we have obtained a chain map between the underlying complexes of $\sL_{\rm YM_2}$ and $\sL_{\rm YM_1}$. In fact, it is a quasi-isomorphism of complexes since this chain map reduces to the identity (modulo constant prefactors) on the cohomologies. 
    
    Moreover, the set of maps $\phi_1$ can be enlarged to include maps $\phi_i:\sL_{\rm YM_2}\times\cdots\times\sL_{\rm YM_2}\to\sL_{\rm YM_1}$ to obtain a fully-fledged quasi-isomorphism~\eqref{eq:L_infty_morphism} between the $L_\infty$-algebras $\sL_{\rm YM_2}$ and $\sL_{\rm YM_1}$. Indeed, the only non-vanishing higher map $\phi_i$ is given by the polarisation of
    \begin{equation}
        \phi_2(A,A)\ \coloneqq\ -\tfrac{1}{\eps}P_+[A,A]~.
    \end{equation}
    In~\cite{Rocek:2017xsj,Jurco:2018sby}, this quasi-isomorphism was given in the $Q$-manifold language which is somewhat more transparent.
    
    Altogether, we conclude that the $L_\infty$-algebra $(\sfL_{\rm YM_2},\mu_i,\langle-,-\rangle_{\sfL_{\rm YM_2}})$ is indeed the strictification of $(\sfL_{\rm YM_1},\mu_i,\langle-,-\rangle_{\sfL_{\rm YM_1}})$.
    
    \subsection{Scattering amplitudes and recursion relations}
    
    \paragraph{Minimal model from the second-order formulation.} The cohomology of the $L_\infty$-algebra~\eqref{eq:ym2_L_algebra} reads as $\sL^\circ_{\rm YM_2}=\sL^\circ_{\rm Maxwell_2}\otimes\mathfrak{su}(N)$ with
    \begin{equation}
        \sL^\circ_{\rm Maxwell_2}\ \coloneqq\ (~\FR\ \xrightarrow{~~~~~}\ \ker(\dd^\dagger \dd)/\im(\dd)\ \xrightarrow{~~~~~}\ \ker(\dd^\dagger \dd)/\im(\dd)\ \xrightarrow{~~~~~}\ \FR ~)~.
    \end{equation}
    We choose the projectors $\sfp_{k}$ to be the evident $L^2$-projectors onto the subspaces $\sL^\circ_{{\rm YM_2},k}\subseteq \sL_{{\rm YM_2},k}$ and we have the trivial embeddings $\sfe_{k}$. To compute the $L_\infty$-structure on $\sL^\circ_{\rm YM_2}$, we need also a contracting homotopy $\sfh=(\sfh_k)$ with $\sfh_k:\sfL_k\to\sfL_{k-1}$ which satisfies~\eqref{eq:contracting_homotopy_htilde}. Some algebra shows that\footnote{See~\cite{Jurco:2018sby} for details on the compact case.}
    \begin{subequations}\label{eq:contracting_homotopy_ym2}
        \begin{equation}
            \sfh_1\ \coloneqq\ G^F\dd^\dagger~,~~~\sfh_2\ \coloneqq\ G^F P_{\rm ex}~,\eand\sfh_3\ \coloneqq\ G^F\dd
        \end{equation}
        is a possible choice. Here, $G^F$ is the Green operator~\eqref{eq:green_operator} and $P_{\rm ex}$ is the projector onto the exact part under the abstract Hodge--Kodaira decomposition as discussed in Section~\ref{ssec:L_infty_algebras} i.e.~onto the image of $\dd^\dagger\dd$. Explicitly, in momentum space and suppressing the gauge algebra for the moment, we have
        \begin{equation}
            \hat \sfh^{\mu\nu}_{2}(k)\ =\ \frac{1}{k^2+\di\eps} \hat P_{\rm ex}^{\mu\nu}(k)~,\ewith  \hat P_{\rm ex}^{\mu\nu}(k)\ =\ \eta^{\mu\nu}-\frac{k^\mu k^\nu}{k^2}~.
        \end{equation}
    \end{subequations}
    Recall that our formulas~\eqref{eq:minimal_model} were derived under the assumption that $\sfh_1(A)=0$, cf.~\eqref{eq:minimal_model_gauge_fixing}. Here, this implies that we work in Lorenz gauge $\dd^\dagger A=0$, and the propagator $G^F P_{\rm ex}$ is indeed the corresponding gluon propagator. 
    
    It remains to insert the projectors and contracting homotopies into~\eqref{eq:minimal_model} to write down the quasi-isomorphism as well as the higher products for the minimal model. 
    
    \paragraph{Berends--Giele gluon recursion relation.}
    Let us denote the generators in the fundamental representation of $\mathfrak{su}(N)$ by $\tau_a$ and set
    \begin{equation}\label{eq:structure_constants}
        [\tau_a,\tau_b]\ =\ {f_{ab}}^c\tau_c\eand g_{ab}\ \coloneqq\ \tr(\tau_a^\dagger \tau_b)\ =\ -\tr(\tau_a \tau_b)\ =\ \tfrac12\delta_{ab}~.
    \end{equation}
    Using $g_{ab}$, we may rewrite the structure constants $f_{abc}\coloneqq{f_{ab}}^d g_{dc}$ as $f_{abc}=-\tr([\tau_a,\tau_b]\tau_c)$. Furthermore, with the help of the completeness relation
    \begin{equation}
        g^{ab}{(\tau_a)_m}^n{(\tau_b)_k}^l\ =\
        -\delta_m^l\delta_k^n+\tfrac{1}{N}\delta_m^n\delta_k^l
    \end{equation}
    we immediately obtain
    \begin{equation}\label{eq:fierz_structure}
        \begin{gathered}
            g^{ab}\tr(X\tau_a)\tr(\tau_bY)\ =\ -\tr(XY)+\tfrac{1}{N}\tr(X)\tr(Y)~,\\
            g^{a_1a_2}g^{b_1b_2}\tr(X\tau_{a_1})\tr(Y\tau_{b_1})f_{a_2b_2c}\ =\ -\tr([X,Y]\tau_c)
        \end{gathered}
    \end{equation}
    for any two matrices $X$ and $Y$. Consequently, all commutators appearing below can be expressed in terms of such traces.
    
    Consider now a plane wave  $A=A_\mu\, \dd x^\mu$ with $A_\mu=\eps_\mu(k)\,\de^{\di k\cdot x}\,X$, where $k_\mu$ is the four-momentum and $\eps_\mu$ the polarisation vector with $k^2= 0$ and $k\cdot\eps=0$, and $X\in\mathfrak{su}(N)$. We shall also write
    \begin{equation}\label{eq:plane_wave_A}
        A(i)\ \coloneqq\ A_\mu(i)\,\dd x^\mu\ewith  A_\mu(i)\ \coloneqq\ \underbrace{\eps_\mu(k_i)}_{=:\,J_\mu(i)}\de^{\di k_i\cdot x}\,X_i~,
    \end{equation}
    to denote the `$i$th gluon'. 
    
    Then, the action of $\phi_1$ in~\eqref{eq:minimal_model} on $A(1)$ is simply given by
    \begin{equation}\label{eq:ym2_phi1}
        \phi_1(A(1))\ =\ \sfe(A(1))\ =\ J_\mu(1)\,\de^{\di k_1\cdot x}X_1\,\dd x^\mu~.
    \end{equation}
    Moreover, the action of $\phi_2$ is
    \begin{subequations}
        \begin{equation}
            \phi_2(A(1),A(2))\ =\ -(\sfh_2\circ\mu_2)(\phi_1(A(1)),\phi_1(A(2)))
        \end{equation}
        and with~\eqref{eq:ym2_phi1} and~\eqref{eq:ym2_brackets}, we find
        \begin{equation}
            \begin{aligned}
                \mu_2(A(1),A(2))\ &=\ \dd^\dagger[A(1),A(2)]+{\star[A(1),{\star\dd A(2)}]}+{\star[A(2),{\star\dd A(1)}]}\\
                &=\ \big\{2(J(1)\cdot k_2)J_\mu(2)-2(J(2)\cdot k_1)J_\mu(1)\,+\\
                &\kern2cm+(J(1)\cdot J(2))(k_1-k_2)_\mu\big\}\de^{\di (k_1+k_2)\cdot x}\,[X_1,X_2]\,\dd x^\mu\\
                &=\ \ldsb J(1),J(2)\rdsb_\mu\, \de^{\di (k_1+k_2)\cdot x}\,[X_1,X_2]\,\dd x^\mu~,
            \end{aligned}
        \end{equation}
        where
        \begin{equation}\label{eq:BG_2_bracket}
            \ldsb J(1),J(2)\rdsb_\mu\ \coloneqq\ 2(J(1)\cdot k_2)J_\mu(2)-2(J(2)\cdot k_1)J_\mu(1)+(J(1)\cdot J(2))(k_1-k_2)_\mu~.
        \end{equation}
        Consequently, using the contracting homotopy~\eqref{eq:contracting_homotopy_ym2}, we obtain
        \begin{equation}\label{eq:ym2_phi2}
            \begin{aligned}
                \phi_2(A(1),A(2))\ &=\  -P_{\rm ex}\left(\frac{\ldsb J(1),J(2)\rdsb_\mu}{(k_1+k_2)^2}\,\de^{\di (k_1+k_2)\cdot x}\,[X_1,X_2]\,\dd x^\mu\right)\\
                \ &=\  -\underbrace{\frac{\ldsb J(1),J(2)\rdsb_\mu}{(k_1+k_2)^2}}_{=:\,J_\mu(1,2)}\,\de^{\di (k_1+k_2)\cdot x}\,[X_1,X_2]\,\dd x^\mu\\
                &=\ -\frac12 \sum_{\sigma\in S_2} J_\mu(\sigma(1),\sigma(2))\,\de^{\di (k_{\sigma(1)}+k_{\sigma(2)})\cdot x}\,[X_{\sigma(1)},X_{\sigma(2)}]\,\dd x^\mu~,
            \end{aligned}
        \end{equation}
    \end{subequations}
    where in the second step, we used that $P_{\rm ex}$ acts trivially and the sum is over all permutations. Equation~\eqref{eq:ym2_phi2} yields indeed the 2-gluon current that can be found in Berends--Giele~\cite{Berends:1987me}. It is also instructive to give the next level expression before turning to the general case. In particular, the action of $\phi_3$ is
    \begin{subequations}
        \begin{equation}
            \begin{aligned}
                \phi_3(A(1),A(2),A(3))\ &=\ -(\sfh_2\circ\mu_2)(\phi_1(A(1)),\phi_2(A(2),A(3)))\,-\\
                &\kern1cm-(\sfh_2\circ\mu_2)(\phi_1(A(2)),\phi_2(A(1),A(3)))\,-\\
                &\kern1cm-(\sfh_2\circ\mu_2)(\phi_1(A(3)),\phi_2(A(1),A(2)))\,-\\
                &\kern1cm-(\sfh_2\circ\mu_3)(\phi_1(A(1)),\phi_1(A(2)),\phi_1(A(3)))~.
            \end{aligned}
        \end{equation}
        From~\eqref{eq:ym2_brackets}, we have
        \begin{equation}
            \begin{aligned}
                & \mu_3(A(1),A(2),A(3))\ =\\
                &\kern.5cm=\ \sum_{\sigma\in C_3}{\star[A(\sigma(1)),\star[A(\sigma(2)),A(\sigma(3))]]}\\
                &\kern.5cm=\  -\sum_{\sigma\in C_3} \ldsb J(\sigma(1)),J(\sigma(2)),J(\sigma(3))\rdsb_\mu\,\de^{\di(k_{\sigma(1)}+k_{\sigma(2)}+k_{\sigma(3)})\cdot x}\, [X_{\sigma(1)},[X_{\sigma(2)},X_{\sigma(3)}]]\,\dd x^\mu~,
            \end{aligned}
        \end{equation}
        where the sum is over cyclic permutations only and
        \begin{equation}\label{eq:part_BG_3_bracket}
            \begin{aligned}
                {\ldsb J(1),J(2),J(3)\rdsb_\mu}\ &\coloneqq\ (J(1)\cdot J(3))J_\mu(2)-(J(1)\cdot J(2))J_\mu(3)~.
            \end{aligned}
        \end{equation}
        Combining this with the expression~\eqref{eq:ym2_phi2} and using the contracting homotopy~\eqref{eq:contracting_homotopy_ym2}, we immediately find that $\phi_3$ is given by
        \begin{equation}\label{eq:ym2_phi3}
            \begin{aligned}
                &\phi_3(A(1),A(2),A(3))\ =\\
                &\kern1cm=\ P_{\rm ex}\sum_{\sigma\in C_3}\tilde J_\mu(\sigma(1),\sigma(2),\sigma(3))\,\de^{\di (k_{\sigma(1)}+k_{\sigma(2)}+k_{\sigma(3)})\cdot x}\,[X_{\sigma(1)},[X_{\sigma(2)},X_{\sigma(3)}]]\,\dd x^\mu~,
            \end{aligned}
        \end{equation}
        where
        \begin{equation}
            \tilde J_\mu(1,2,3)\ \coloneqq\ \frac{\ldsb J(1),J(2,3)\rdsb_\mu+\ldsb J(1),J(2),J(3)\rdsb_\mu}{(k_1+k_2+k_3)^2}~.
        \end{equation}
        The expression for the 3-gluon current as given by Berends--Giele~\cite{Berends:1987me} is simply
        \begin{equation}
            J_\mu(1,2,3)\ \coloneqq\ \tilde J_\mu(1,2,3)-\tilde J_\mu(3,1,2)~,
        \end{equation}
        and, upon using the antisymmetry and the Jacobi identity for the Lie bracket $[-,-]$, a short calculation reveals that~\eqref{eq:ym2_phi3} becomes
        \begin{equation}
            \begin{aligned}
                &\phi_3(A(1),A(2),A(3))\ =\\
                &\kern1cm=\ \frac{1}{3}\sum_{\sigma\in S_3} J_\mu(\sigma(1),\sigma(2),\sigma(3))\,\de^{\di (k_{\sigma(1)}+k_{\sigma(2)}+k_{\sigma(3)})\cdot x}\,[X_{\sigma(1)},[X_{\sigma(2)},X_{\sigma(3)}]]\,\dd x^\mu~,
            \end{aligned}
        \end{equation}
    \end{subequations}
    where the sum here is over all permutations and $P_{\rm ex}$ acts again trivially. 
    
    Let us now turn to the general case. The above discussion for 2- and 3-points motivates us to define
    \begin{equation}
        J_a(1,\ldots,i)\ =\ g_{ab} J^b(1,\ldots,i)\ \coloneqq\  -\tr(\phi_i(A(1),\ldots, A(i))\tau_a)
    \end{equation}
    with $g_{ab}$ as given in~\eqref{eq:structure_constants}. Hence,
    \begin{equation}
        \phi_i(A(1),\ldots, A(i))\ =\ J^a(1,\ldots,i)\tau_a~.
    \end{equation}
    Furthermore, we also define
    \begin{equation}\label{eq:def_hat_j}
        \begin{gathered}
            J^a(1,\ldots,i)\ =:\ g^{ab}\sum_{\sigma\in S_i}\tr(X_{\sigma(1)}\cdots X_{\sigma(i)}\tau_b)J_\mu (\sigma(1),\ldots,\sigma(i))\,\de^{\di(k_{\sigma(1)}+\cdots+k_{\sigma(i)})\cdot x}\dd x^\mu\\
            J(1,\ldots,i)\ \coloneqq\ J_\mu(1,\ldots,i)\,\dd x^\mu
        \end{gathered}
    \end{equation}
    similar to Berends--Giele~\cite{Berends:1987me}. Then, the first term in the quasi-isomorphism
    \begin{equation}\label{eq:phi_min_mod_BG}
        \begin{aligned}
            &\phi_i(A(1),\ldots,A(i))\ =\\
            &=\ -\frac{1}{2!} \sum_{k_1+k_2=i}\sum_{\sigma\in{\rm Sh}(k_1;i)}\times\\
            &\kern1.5cm\times (\sfh_2\circ\mu_2)\big(\phi_{k_1}\big(A(\sigma(1),\ldots,A(\sigma(k_1))\big),\phi_{k_2}\big(A(\sigma(k_1+1),\ldots,A(\sigma(i))\big)\big)\,-\\
            &\kern.55cm -\frac{1}{3!} \sum_{k_1+k_2+k_3=i}\sum_{\sigma\in{\rm Sh}(k_1,k_2;i)}\times\\
            &\kern1.5cm\times (\sfh_2\circ\mu_3)\big(\phi_{k_1}\big(A(\sigma(1),\ldots,A(\sigma(k_1))\big),\ldots,\phi_{k_3}\big(A(\sigma(k_1+k_2+1),\ldots,A(\sigma(i))\big)\big)
        \end{aligned}
    \end{equation}
    is given by
    \begin{equation}
        \begin{aligned}
            {\rm (I)}\ &\coloneqq\  -\frac{1}{2!} \sum_{k_1+k_2=i}\sum_{\sigma\in{\rm Sh}(k_1;i)}\times\\
            &\kern1.5cm\times \mu_2\big(\phi_{k_1}\big(A(\sigma(1),\ldots,A(\sigma(k_1))\big),\phi_{k_2}\big(A(\sigma(k_1+1),\ldots,A(\sigma(i))\big)\big)\\
            &\phantom{:}=\ -\frac{1}{2!} \sum_{\sigma\in S_i}\sum_{j=1}^{i-1}\frac{1}{j!(i-j)!}\times\\
            &\kern1.5cm\times \mu_2\big(\phi_{j}\big(A(\sigma(1),\ldots,A(\sigma(j))\big),\phi_{i-j}\big(A(\sigma(j+1),\ldots,A(\sigma(i))\big)\big)\\
            &\phantom{:}=\ -\frac{1}{2!}\sum_{\sigma\in S_i}\sum_{j=1}^{i-1}\frac{1}{j!(i-j)!}\ldsb J^a(\sigma(1),\ldots,\sigma(j)), J^b(\sigma(j+1),\ldots,\sigma(i))\rdsb f_{abc}g^{cd}\tau_d\\
            &\phantom{:}=\ \sum_{\sigma\in S_i}\sum_{j=1}^{i-1}\ldsb J(\sigma(1),\ldots,\sigma(j)),J(\sigma(j+1),\ldots,\sigma(i))\rdsb\,\times\\
            &\kern1.5cm\times \de^{\di(k_{\sigma(1)}+\cdots+k_{\sigma(i)})\cdot x}\,g^{ab}\tr(X_{\sigma(1)}\cdots X_{\sigma(i)}\tau_b)\tau_a~,
        \end{aligned}
    \end{equation}
    where we have substituted~\eqref{eq:def_hat_j} and used~\eqref{eq:fierz_structure}. In addition, $\ldsb-,-\rdsb$ is the bracket defined in~\eqref{eq:BG_2_bracket}.
    
    Likewise, the second term in~\eqref{eq:phi_min_mod_BG} is given by
    \begin{equation}
        \begin{aligned}
            {\rm (II)}\ &\coloneqq\   -\frac{1}{3!} \sum_{k_1+k_2+k_3=i}\sum_{\sigma\in{\rm Sh}(k_1,k_2;i)}\times\\
            &\kern1.5cm\times \mu_3\big(\phi_{k_1}\big(A(\sigma(1),\ldots,A(\sigma(k_1))\big),\ldots,\phi_{k_3}\big(A(\sigma(k_1+k_2+1),\ldots,A(\sigma(i))\big)\big)\\
            &\phantom{:}=\ -\frac{1}{3!} \sum_{\sigma\in S_i}\sum_{j=1}^{i-2}\sum_{k=j+1}^{i-1}\frac{1}{j!(k-j)!(i-k)!}\mu_3\big(\phi_{j}\big(A(\sigma(1),\ldots,A(\sigma(j))\big),\\
            &\kern3cm \phi_{k-j}\big(A(\sigma(j+1),\ldots,A(\sigma(k))\big),\phi_{i-k}\big(A(\sigma(k+1),\ldots,A(\sigma(i))\big)\big)\\
            &\phantom{:}=\ \frac{1}{2!} \sum_{\sigma\in S_i}\sum_{j=1}^{i-2}\sum_{k=j+1}^{i-1}\frac{1}{j!(k-j)!(i-k)!}\times\\
            &\kern1.5cm\times \ldsb J^a(\sigma(1),\ldots,\sigma(j)), J^b(\sigma(j+1),\ldots,\sigma(k)),J^c(\sigma(k+1),\ldots,\sigma(i))\rdsb \,\times\\
            &\kern1.5cm\times f_{bcd}f_{aef} g^{de}g^{fg}\tau_g\\
            &\phantom{:}=\ \sum_{\sigma\in S_i}\sum_{j=1}^{i-2}\sum_{k=j+1}^{i-1} \ldsb J(\sigma(1),\ldots,\sigma(j)),J(\sigma(j+1),\ldots,\sigma(k)),J(\sigma(k+1),\ldots,\sigma(i))\rdsb' \,\times\\
            &\kern1.5cm\times \de^{\di(k_{\sigma(1)}+\cdots+k_{\sigma(i)})\cdot x}\,g^{ab}\tr(X_{\sigma(1)}\cdots X_{\sigma(i)}\tau_b)\tau_a~,
        \end{aligned}
    \end{equation}
    where we have again substituted~\eqref{eq:def_hat_j}, used twice the relations~\eqref{eq:fierz_structure}, and defined
    \begin{equation}
        {\ldsb J(1),J(2),J(3)\rdsb}'\ \coloneqq\ \ldsb J(1),J(2),J(3)\rdsb-\ldsb J(3),J(1),J(2)\rdsb
    \end{equation}
    with $\ldsb-,-,-\rdsb$ the bracket introduced in~\eqref{eq:part_BG_3_bracket}. Hence, upon adding (I) and (II) and applying the contracting homotopy $\sfh_2$ from~\eqref{eq:contracting_homotopy_ym2}, we find 
    \begin{equation}\label{eq:i_colour_stripped_gluon_current}
        \begin{aligned}
            &J(1,\ldots,i)\ =\\
            &\kern.5cm=\ \frac{1}{(k_{1}+\cdots+k_{i})^2}\,\times\\
            &\kern1.5cm\times \hat P_{\rm ex}\left\{\sum_{j=1}^{i-1}\ldsb J(1,\ldots,j),J(j+1,\ldots,i)\rdsb\,+\right. \\
            &\kern3.5cm\left.+\sum_{j=1}^{i-2}\sum_{k=j+1}^{i-1}\ldsb J(1,\ldots,j),J(j+1,\ldots,k),J(k+1,\ldots,i)\rdsb'\right\}.
        \end{aligned}
    \end{equation}
    This is precisely the Berends--Giele recursion~\cite{Berends:1987me} modulo the appearance of the projector $\hat P_{\rm ex}$. As before, it acts trivially, as follows from the current conservation property of the expression inside the curly bracket, that is, $(k_1+\cdots+k_i)\cdot\{\cdots\}=0$.
    
    Altogether, we conclude that the quasi-isomorphism between the $L_\infty$-algebra governing Yang--Mills theory in the second-order formulation and its minimal model encodes the Berends--Giele gluon current recursion relations. The actual scattering amplitudes $\CA(1,\dots,i)$  now follow directly from the homotopy Maurer--Cartan action~\eqref{eq:hMC_action} for the minimal model brackets~\eqref{eq:minimal_model} for this quasi-isomorphism. For $i\geq2$, we have
    \begin{subequations}
        \begin{equation}\label{eq:amplitude_ym2}
            \CA(1,\dots,i+1)\ =\ \langle A(1),\mu_i^\circ(A(2),\ldots,A(i+1))\rangle_{\sfL_{\rm YM_2}}
        \end{equation}
        with
        \begin{equation}
            \begin{aligned}
                &\mu_i^\circ(A(1),\ldots,A(i))\ =\\
                &\kern1cm=\ -\sum_{\sigma \in S_i}(k_1+\cdots+k_i)^2\,J_\mu(\sigma(1),\ldots,\sigma(i))\,\de^{\di (k_{\sigma(1)}+\cdots+k_{\sigma(i)})\cdot x}\,\times\\
                &\kern3cm\times g^{ab}\tr(X_{\sigma(1)}\cdots X_{\sigma(i)}\tau_b)\tau_a\,\dd x^\mu\Big|_{(k_1+\cdots+k_i)^2=0}~,
            \end{aligned}
        \end{equation}
    \end{subequations}
    where $J_\mu(1,\ldots,i)$ as given in~\eqref{eq:i_colour_stripped_gluon_current}. Note that the expression $\mu_i^\circ(A(1),\ldots,A(i))$ is already co-closed and hence, the projection $\sfp$ in~\eqref{eq:minimal_model} acts by requiring that $(k_1+\cdots+k_i)^2=0$ in the case at hand. Note also that the symmetry of the amplitude~\eqref{eq:amplitude_ym2} under the exchange of any two gluons is due to the cyclicity~\eqref{eq:cyclicity} of the inner product~\eqref{eq:inner_product_YM2}.

    \paragraph{Minimal model from the strictification.} 
    We could also have constructed a minimal model and corresponding recursion relations for tree-level scattering amplitudes from the strictified $L_\infty$-algebra $(\sfL_{\rm YM_1},\mu_i,\langle-,-\rangle_{\sfL_{\rm YM_1}})$. See~\cite{Jurco:2018sby} for the construction of the contracting homotopy in this case. Any resulting minimal model $\sfL^\circ_{\rm YM_1}$ is certainly $L_\infty$-isomorphic to $\sfL^\circ_{\rm YM_2}$ but the shape of the recursion relation is particularly suited for discussing the BCFW recursion relations~\cite{Britto:2004ap,Britto:2005fq} as shown in~\cite{Nutzi:2018vkl}, because only trivalent vertices are present in $(\sfL_{\rm YM_1},\mu_i,\langle-,-\rangle_{\sfL_{\rm YM_1}})$. In addition, this also simplifies the off-shell recursion relations~\eqref{eq:i_colour_stripped_gluon_current}.
    
    \datamanagement
    
    No additional research data beyond the data presented and cited in this work are needed to validate the research findings in this work.
    
    \acknowledgements 
    
    We would like to thank Alexandros Arvanitakis, Bin Cheng, Heiko Gimperlein, Andrea Prinsloo, Alexander Schenkel, Paul Skerritt, and Alessandro Torrielli for helpful discussions. We are particularly grateful to Andrea N{\"u}tzi and Michael Reiterer for asking questions that led to corrections and improvements of the discussion in Section~\ref{sec:scalar_FT}. T.M.~was partially supported by the EPSRC grant EP/N509772. 
    
    \appendices
    
    \subsection{Proof of the minimal model recursion relations} \label{app:proof_minimal_model}
    
    To derive the recursion relations~\eqref{eq:minimal_model}, we need to construct a quasi-isomorphism $\phi:\sL^\circ\rightarrow\sL$ that allows us to pull back the higher products on $\sL$ to $\sL^\circ$ via formula~\eqref{eq:L_infty_morphism}. Our construction of $\phi$ follows the idea of~\cite{Kajiura:0306332}, where essentially the same construction was given in the case of $A_\infty$-algebras. In particular, we assume that we have a Maurer--Cartan element $a^\circ$ in $\sL^\circ$ and map it to an element $a$ in $\sL$. The fact that Maurer--Cartan elements are mapped to Maurer--Cartan elements under quasi-isomorphisms, cf.~\eqref{eq:map_hMC_to_hMC}, together with the assumption that $a^\circ$ (and therefore $a$) is small, will give us enough constraints to determine the quasi-isomorphisms and the higher products on $\sL^\circ$.
    
    \paragraph{Proof.} We start from the contracting homotopy 
    \begin{equation}
        \myxymatrix{\ar@(dl,ul)[]^h \sfL~\ar@<+2pt>@{->>}[rr]^{\kern-20pt \sfp} & & ~H^\bullet_{\mu_1}(\sfL) \ar@<+2pt>@{^(->}[ll]^{\kern-20pt \sfe}},
    \end{equation}
    where we can assume that $\sfh^2=0$ and $\sfe\circ \sfp$, $\mu_1\circ \sfh$ and $\sfh\circ \mu_1$ are projectors onto $\sL_{\rm harm}$, $\sL_{\rm ex}$, and $\sL_{\rm coex}$, respectively. Moreover, let $a^\circ\in \sL^\circ_1$ be a Maurer--Cartan element. Under a quasi-isomorphism $\phi$, $a^\circ$ is mapped to
    \begin{equation}\label{eq:hMCelt_a_from_a_circ}
        a\ =\ \sum_{i\geq 1}\frac{1}{i!}\phi_i(a^\circ,\dots,a^\circ)~.
    \end{equation}
    A convenient choice is $\phi_1=\sfe$, and it remains to identify $\phi_i$ for $i>1$. We will do this by fixing $a$ as a function of $a^\circ$. 
    
    Recall that~\eqref{eq:Hodge-Kodaira-decomposition} yields the unique decomposition
    \begin{equation}
        a\ =\ a_{\rm harm}+a_{\rm ex}+a_{\rm coex}~,\ewith a_{\rm harm,\,ex,\,coex}\ \in\ \sL_{\rm harm,\,ex,\,coex}~.
    \end{equation}
    There is some freedom in the choice of $\phi$ and without loss of generality, we may impose the \emph{gauge fixing} condition 
    \begin{equation}\label{eq:minimal_model_gauge_fixing}
        \sfh(a)\ =\ 0~.
    \end{equation} 
    This is, in fact, a generalisation of the Lorenz gauge fixing condition from ordinary gauge theory. Consequently, $a_{\rm ex}=(\mu_1\circ\sfh)(a)=0$. Moreover, the fact that $\mu_1$ is a chain map implies that $\mu_1(a_{\rm harm})=(\mu_1\circ \sfe\circ \sfp)(a)=0$ so that the homotopy Maurer--Cartan equation for $a$ becomes
    \begin{equation}\label{eq:hmc_coex}
        \mu_1(a_{\rm coex})+\sum_{i\geq 2}\frac{1}{i!}\mu_i(a_{\rm harm}+a_{\rm coex},\ldots,a_{\rm harm}+a_{\rm coex})\ =\ 0~.
    \end{equation}
    Upon acting with $\sfh$ on both sides of this equation, we obtain 
    \begin{equation}\label{eq:algebraic_relation_ac}
        a_{\rm coex}\ =\ -\sum_{i\geq 2}\frac{1}{i!}(\sfh\circ\mu_i)(a_{\rm harm}+a_{\rm coex},\ldots,a_{\rm harm}+a_{\rm coex})~.
    \end{equation}
    
    If we now assume that $a^\circ$ is small, say of order $\CO(g)$ with $g \ll 1$ for $g$ a formal parameter, we may rewrite~\eqref{eq:hMCelt_a_from_a_circ} as
    \begin{subequations}\label{eq:hbar_expansion_a}
        \begin{equation}
            \begin{aligned}
                a\ &=\ \sum_{i\geq 1}\frac{g^i}{i!}\phi_i(a^\circ,\dots,a^\circ)\ =\ g\underbrace{\sfe(a^\circ)}_{=:\,a^{(1)}}+\frac{g^2}{2}\underbrace{\phi_2(a^\circ,a^\circ)}_{=:\,a^{(2)}}+\cdots\\
                &=\ g\big(a^{(1)}_{\rm harm}+a^{(1)}_{\rm coex}\big)+\frac{g^2}{2}\big(a^{(2)}_{\rm harm}+a^{(2)}_{\rm coex}\big)+\cdots\\
            \end{aligned}
        \end{equation}
        We can then compute the solution $a$ of the homotopy Maurer--Cartan equation order by order in $g$ using~\eqref{eq:algebraic_relation_ac}. In this process, we can choose to put $a^{(i)}_{\rm harm}=0$ for $i>1$ so that
        \begin{equation}
            \begin{aligned}
                a\ &=\ \underbrace{g\, a^{(1)}_{\rm harm}}_{=\,a_{\rm harm}} +\underbrace{\sum_{i\geq 2}\frac{g^i}{i!} a^{(i)}_{\rm coex}}_{=\,a_{\rm coex}}\ =\ a_{\rm harm}+a_{\rm coex}~.
            \end{aligned}
        \end{equation}
    \end{subequations}
    Substituting this expansion into~\eqref{eq:algebraic_relation_ac}, we arrive at the recursion relation
    \begin{equation}\label{eq:recursion_relation_ac}
        a_{\rm coex}^{(i)}\ =\ -\sum_{j= 2}^i\frac{1}{j!}\sum_{k_1+\cdots+k_j=i}(\sfh\circ\mu_j)(a_{\rm harm}^{(k_1)}+a_{\rm coex}^{(k_1)},\ldots,a_{\rm harm}^{(k_j)}+a_{\rm coex}^{(k_j)})
    \end{equation}
    for $a_{\rm coex}$. Comparison with~\eqref{eq:hMCelt_a_from_a_circ} then yields the quasi-isomorphism~\eqref{eq:minimal_model} when evaluated at degree~$1$ elements.
    
    To recover also the brackets $\mu_i^\circ$ on $\sL^\circ$ listed in~\eqref{eq:minimal_model} by pullback, we note that upon applying the projector $\sfp$ to~\eqref{eq:hmc_coex} and using the fact that $\sfp$ is a chain map, we immediately find that
    \begin{equation}
        \sum_{i\geq 2}\frac{1}{i!}(\sfp\circ\mu_i)(a_{\rm harm}+a_{\rm coex},\ldots,a_{\rm harm}+a_{\rm coex})\ =\ 0~.
    \end{equation}
    Hence, after substituting the expansion~\eqref{eq:hbar_expansion_a}, we recover the brackets~\eqref{eq:minimal_model} for degree~$1$ elements.
    
    Our derivation above is strictly speaking only applicable to Maurer--Cartan elements, which are elements of the $L_\infty$-algebra of degree~1. As noted in~\cite{Jurco:2018sby}, however, we may enlarge every $L_\infty$-algebra $\sL$ to the $L_\infty$-algebra $\sL_{\CCC}\coloneqq\CCC^\infty(\sL[1])\otimes \sL$ where $\CCC^\infty(\sL[1])$ are the smooth functions on the grade-shifted vector space $\sL[1]$. Then, every element in $\sL$ gives rise to a degree~$1$ element in $\sL_{\CCC}$, and, applying the above construction to $\sL_{\CCC}$ yields the full $L_\infty$-quasi-isomorphism and brackets listed in~\eqref{eq:minimal_model}.
    
    \paragraph{\mathversion{bold}Cyclic $L_\infty$-algebras.}
    Finally, we note that the above construction also extends to the cyclic case. For this, we need $\sfh$ chosen such that 
    \begin{equation}\label{eq:coexcoex0}
        \langle \sL_{\rm coex},\sL_{\rm coex}\rangle_{\sL}\ =\ 0~.
    \end{equation}
    This is always possible since cyclicity~\eqref{eq:cyclicity} for $\mu_1$ implies in general that
    \begin{equation}
        \langle \sL_{\rm ex},\sL_{\rm ex}\rangle_{\sL}\ =\ \langle \sL_{\rm harm},\sL_{\rm ex}\rangle_{\sL}\ =\ 0~.
    \end{equation}
    The remaining freedom in the choice of $\sfh$ can therefore be used to ensure that the only non-vanishing entries of the underlying metric are 
    \begin{equation}
        \langle \sL_{\rm harm},\sL_{\rm harm}\rangle_{\sL}~,~~~\langle \sL_{\rm ex},\sL_{\rm coex}\rangle_{\sL}~,\eand \langle \sL_{\rm coex},\sL_{\rm ex}\rangle_{\sL}~.
    \end{equation}
    
    If we now pull-back the cyclic structure from $\sL$ to $\sL^\circ$ and define
    \begin{equation}
        \langle \ell_1^\circ,\ell_2^\circ\rangle_{\sfL^\circ}\ \coloneqq\ \langle \phi_1(\ell^\circ_1),\phi_1(\ell^\circ_2)\rangle_{\sfL}~,
    \end{equation}
    we have satisfied the first condition in~\eqref{eq:compatibility_cyclic_morphism} on a morphism of cyclic $L_\infty$-algebras. The second condition in~\eqref{eq:compatibility_cyclic_morphism} is implied by~\eqref{eq:coexcoex0} together with $\im(\phi)\subseteq \sL_{\rm coex}$.
    
    \subsection{Dynkin--Specht--Wever lemma}
    
    \paragraph{Statement.}
    For simplicity, let $\fra$ be a matrix algebra and $\frl$ be the Lie subalgebra generated by the elements that generate $\fra$, that is, the \emph{free Lie algebra} over $\fra$. Consider the \emph{Dynkin map} $D:\fra\to\frl$ defined by
    \begin{equation}
        \fra\ \ni\ \sum_{\sigma\in S_i}\lambda_\sigma\, X_{\sigma(1)}\cdots X_{\sigma(i)}\ \mapsto\ \sum_{\sigma\in S_i}\lambda_\sigma\, [X_{\sigma(1)},[X_{\sigma(2)},\ldots[X_{\sigma(i-1)}, X_{\sigma(i)}]\cdots]]\ \in\ \frl~,
    \end{equation}
    where $X_1,\ldots,X_i\in\fra$ and the coefficients $\lambda_\sigma$ are some numbers. The \emph{Dynkin--Specht--Wever lemma} then asserts that if $p(X)\coloneqq\sum_{\sigma\in S_{i_p}}\lambda_\sigma X_{\sigma(1)}\cdots X_{\sigma(i_p)}\in\frl$ then
    \begin{equation}\label{eq:DSW_lemma}
        D(p(X))\ =\ i_p \,p(X)~.
    \end{equation}
    Hence, for any homogeneous polynomial $p(X)\in\fra$ of degree~$i_p$, we obtain $(D\circ D)(p(X))=i_p D(p(X))$.
    
    \paragraph{Proof.}
    To prove~\eqref{eq:DSW_lemma}, we follow~\cite{lyndon1955}. Firstly, we set $\ad(X)(Y)\coloneqq[X,Y]$. Then, one can show by induction on the degree of the polynomial $p(X)$ that if $p(X)\in\frl$ then 
    \begin{subequations}
        \begin{equation}\label{eq:ad_lie}
            \ad(p(X))\ =\ p(\ad(X))
        \end{equation}
        with
        \begin{equation}
            p(\ad(X))\ \coloneqq\ \sum_{\sigma\in S_{i_p}}\lambda_\sigma^{(p)}\, \ad(X_{\sigma(1)})\circ\cdots \circ\ad(X_{\sigma(i_p)})~.
        \end{equation}
    \end{subequations}
    Secondly, \eqref{eq:DSW_lemma} is certainly true for $i_p=1$ so let us assume it is true for $i_p>1$ and prove the statement by induction. To this end, let $p(X)\in\frl$ and $q(X)\in\frl$ be homogeneous polynomials of degrees~$i_p$ and $i_q$, respectively. Then,
    \begin{equation}
        \begin{aligned}
            D(p(X)q(X))\ &=\ \sum_{\sigma\in S_{i_p}}\lambda_\sigma^{(p)} [X_{\sigma(1)},[X_{\sigma(2)},\ldots[X_{\sigma(i_p-1)}, [X_{\sigma(i_p)},D(q(X))]]\cdots]]\\
            &=\ p(\ad(X))(D(q(X))\\
            &=\ \ad(p(X))(D(q(X))\\
            &=\ [p(X),D(q(X))]\\
            &=\ i_q [p(X),q(X)]~,
        \end{aligned}
    \end{equation}
    where in the third step we have used~\eqref{eq:ad_lie} since $q(X)\in\frl$ and in the fifth step the induction hypothesis. Thus,
    \begin{equation}\label{eq:induction_dsw_lemma}
        D([p(X),q(X)])\ =\ (i_p+i_q)[p(X),q(X)]~.
    \end{equation}
    This concludes the proof of~\eqref{eq:DSW_lemma}.
    
    \paragraph{Applications.}
    Consider now
    \begin{equation}\label{eq:dynkin_nested_commutator}
        \begin{aligned}
            D(X_1\cdots X_i)\ &=\ [X_1,[X_{2},\ldots[X_{i-1},X_i]\cdots]]\\
            &=\ \sum_{j=0}^{i-1}\sum_{\sigma\in{\rm Sh}(j;i-1)}(-1)^{i+j+1} X_{\sigma(1)}\cdots X_{\sigma(j)}X_i X_{\sigma(i-1)}\cdots X_{\sigma(j+1)}\\
            &=\ \frac{1}{i} \sum_{j=0}^{i-1}\sum_{\sigma\in{\rm Sh}(j;i-1)}(-1)^{i+j+1}  D(X_{\sigma(1)}\cdots X_{\sigma(j)}X_i X_{\sigma(i-1)}\cdots X_{\sigma(j+1)})~,
        \end{aligned}
    \end{equation}
    where in the third step we have used~\eqref{eq:DSW_lemma}.
    
    Then, again using~\eqref{eq:DSW_lemma}, we obtain
    \begin{equation}\label{eq:nested_2_commutator}
        \begin{aligned}
            \underbrace{[D(X_{1}\cdots X_{i}),D(X_{i+1}\cdots X_{i+j})]}_{=:\,(i+j)\sum_{\sigma\in S_{i+j}}\lambda_\sigma^{(i;i+j)} X_{\sigma(1)}\cdots X_{\sigma(i+j)}}\ &=\ \frac{1}{i+j} D([D(X_{1}\cdots X_{i}),D(X_{i+1}\cdots X_{i+j})])\\
            &=\ \sum_{\sigma\in S_{i+j}}\lambda_\sigma^{(i;i+j)} D(X_{\sigma(1)}\cdots X_{\sigma(i+j)})~,
        \end{aligned}
    \end{equation}
    where the $\lambda_\sigma^{(i;i+j)}$ are given in terms of the coefficients in~\eqref{eq:dynkin_nested_commutator}.
    
    Likewise, again using~\eqref{eq:DSW_lemma}, we have
    \begin{subequations}\label{eq:nested_3_commutator}
        \begin{equation}
            \begin{aligned}
                &{[D(X_{1}\cdots X_{i}),[D(X_{i+1}\cdots X_{i+j}),D(X_{i+j+1}\cdots X_{i+j+k})]]}\ =\\
                &\kern.5cm=\ \frac{1}{(j+k)(i+j+k)}D([D(X_{1}\cdots X_{i}),D([D(X_{i+1}\cdots X_{i+j}),D(X_{i+j+1}\cdots X_{i+j+k})])])\\
                &\kern.5cm=\ \frac{1}{(i+j+k)}\sum_{\sigma_2\in S_{j+k}}\lambda^{(j;j+k)}_{\sigma_2}D([D(X_{1}\cdots X_{i}),D(X_{i+\sigma_2(1)}\cdots X_{i+\sigma_2(j+k)})])])\\
                &\kern.5cm=\ \sum_{\substack{\sigma_1\in S_{i+j+k}\\ \sigma_2\in S_{j+k}}}\lambda^{(i;i+j+k)}_{\sigma_1}\lambda^{(j;j+k)}_{\sigma_2}D(X_{\sigma_1(1)}\cdots X_{\sigma_1(i)}X_{\sigma_1(i+\sigma_2(1))}\cdots X_{\sigma_1(i+\sigma_2(j+k))})\\
                &\kern.5cm=:\ \sum_{\sigma\in S_{i+j+k}}\lambda^{(i,j;i+j+k)}_\sigma D(X_{\sigma(1)}\cdots X_{\sigma(i+j+k)})~,
            \end{aligned}
        \end{equation}
        where the coefficients $\lambda^{(i,j)}_\sigma$ are defined as follows: letting
        \begin{equation}
            \sigma_3\ \coloneqq\ \sigma_1\circ\tau_{\sigma_2}~,\ewith \tau_{\sigma_2}(\ell)\ \coloneqq\ \begin{cases} \ell  &\efor \ell\in\{1,\ldots,i\}~,\\ i+\sigma_2(\ell-i) &\efor \ell\in\{i+1,\ldots,i+j+k\}~,\end{cases}
        \end{equation}
        we obtain
        \begin{equation}
            \begin{aligned}
                &\sum_{\sigma_1\in S_{i+j+k}}\sum_{\sigma_2\in S_{j+k}}\lambda^{(i;i+j+k)}_{\sigma_1}\lambda^{(j;j+k)}_{\sigma_2}D(X_{\sigma_1(1)}\cdots X_{\sigma_1(i)}X_{\sigma_1(i+\sigma_2(1))}\cdots X_{\sigma_1(i+\sigma_2(j+k))})\ =\\
                &\kern.5cm=\ \sum_{\sigma_3\in S_{i+j+k}}\sum_{\sigma_2\in S_{j+k}}\lambda^{(i;i+j+k)}_{\sigma_3\circ\tau^{-1}_{\sigma_2}}\lambda^{(j;j+k)}_{\sigma_2} D(X_{\sigma_3(1)}\cdots X_{\sigma_3(i+j+k)})~,
            \end{aligned}
        \end{equation}
        since when $\sigma_1$ runs over all of $S_{i+j+k}$ so does $\sigma_3$. Consequently, we may set
        \begin{equation}
            \lambda_\sigma^{(i,j;i+j+k)}\ \coloneqq\ \sum_{\sigma'\in S_{j+k}}\lambda^{(i;i+j+k)}_{\sigma\circ\tau^{-1}_{\sigma'}}\lambda^{(j;j+k)}_{\sigma'}~.
        \end{equation}
    \end{subequations}
    
    \subsection{Gluon recursion for general Lie groups}\label{app:BG_compact_simple}
    
    Let us present a derivation of the Berends--Giele recursion from the quasi-isomor-\linebreak phism~\eqref{eq:phi_min_mod_BG} in the case of a general gauge group not necessarily simple and compact, and which uses the Dynkin--Specht--Wever lemma discussed in the previous section.
    
    We again consider plane waves of the form~\eqref{eq:plane_wave_A} and make the ansatz
    \begin{equation}
        \begin{aligned}
            \phi_i(A(1),\ldots,A(i))\ &=\ -\frac{(-1)^i}{i}\sum_{\sigma \in S_i}J_\mu(\sigma(1),\ldots,\sigma(i))\,\de^{\di (k_{\sigma(1)}+\cdots+k_{\sigma(i)})\cdot x}\,\times\\
            &\kern1.5cm\times [X_{\sigma(1)},[X_{\sigma(2)},[\ldots,[X_{\sigma(i-2)},[X_{\sigma(i-1)},X_{\sigma(i)}]]\cdots]]\,\dd x^\mu~.
        \end{aligned}
    \end{equation}
    Upon substituting this into~\eqref{eq:phi_min_mod_BG} and using the contracting homotopy~\eqref{eq:contracting_homotopy_ym2}, a straightforward calculation shows that
    \begin{subequations}
        \begin{equation}
            \begin{aligned}
                &J_\mu(1,\ldots,i)\ =\\
                &\kern.5cm=\ \frac{1}{(k_{1}+\cdots+k_{i})^2}\,\times\\
                &\kern1.5cm\times P_{\rm ex}\left\{\sum_{j=1}^{i-1}\ldsb J(1,\ldots,j),J(j+1,\ldots,i)\rdsb'_\mu\,+\right. \\
                &\kern3.5cm\left.+\sum_{j=1}^{i-2}\sum_{k=j+1}^{i-1}\ldsb J(1,\ldots,j),J(j+1,\ldots,k),J(k+1,\ldots,i)\rdsb''_\mu\right\}
            \end{aligned}
        \end{equation}
        with
        \begin{equation}
            \begin{aligned}
                & {\ldsb J(1,\ldots,j)),J(j+1,\ldots,i)\rdsb'_\mu}\ \coloneqq\\
                &\kern.5cm\coloneqq\ \frac{i}{2j(i-j)}\sum_{\sigma\in S_i}\lambda^{(j;i)}_{\sigma^{-1}}\ldsb J(\sigma(1),\ldots,\sigma(j)),J(\sigma(j+1),\ldots,\sigma(i))\rdsb_\mu~,\\
                &\ldsb J(1,\ldots,j),J(j+1,\ldots,k),J(k+1,\ldots,i)\rdsb'_\mu\ \coloneqq\\
                &\kern.5cm\coloneqq\  \frac{i}{3j(k-j)(i-k)}\sum_{\sigma\in S_i}\lambda^{(j,k-j;i)}_{\sigma^{-1}}\times\\
                &\kern2cm\times\ldsb J(\sigma(1),\ldots,\sigma(j)),J(\sigma(j+1),\ldots,\sigma(k)),J(\sigma(k+1),\ldots,\sigma(i))\rdsb_\mu~,\\
                &\ldsb J(1),J(2),J(3)\rdsb''_\mu\ \coloneqq\ \ldsb J(1),J(2),J(3)\rdsb'_\mu-\ldsb J(3),J(1),J(2)\rdsb'_\mu~,
            \end{aligned}
        \end{equation}
    \end{subequations}
    where $\ldsb-,-\rdsb_\mu$ and $\ldsb-,-,-\rdsb_\mu$ were introduced in~\eqref{eq:BG_2_bracket} and~\eqref{eq:part_BG_3_bracket} and the $\lambda$-coefficients are defined in~\eqref{eq:nested_2_commutator} and~\eqref{eq:nested_3_commutator}, respectively. This is the Berends--Giele recursion for any gauge algebra.
    
    \bibliographystyle{latexeu}
    \bibliography{bigone2}
    
\end{document}